\documentclass[conference,compsoc]{IEEEtran}

\usepackage{xspace}
\usepackage{csquotes}
\usepackage{xcolor}
\usepackage{amsmath}
\usepackage{subcaption}
\usepackage{hyperref}
\usepackage[final]{changes}
\usepackage[ruled,vlined]{algorithm2e}
\usepackage[nolist,nohyperlinks]{acronym}
\usepackage{adjustbox}
\graphicspath{{figures/}}

\acrodefplural{AS}[ASes]{Autonomous Systems}

\newcommand{\etal}{et al.\xspace}
\newcommand{\eg}{e.g.\xspace}
\newcommand{\ie}{i.e.\xspace}
\newcommand{\vfour}{IPv4\xspace}
\newcommand{\vsix}{IPv6\xspace}
\newcommand{\eui}{EUI-64\xspace}
\newcommand{\wifi}{WiFi\xspace}
\newcommand{\wan}{\ac{WAN}\xspace}
\newcommand{\bt}{Bluetooth\xspace}
\newcommand{\wigle}{WiGLE\xspace}
\newcommand{\cpe}{\ac{CPE}\xspace}
\newcommand{\wtb}{WAN-to-BSSID\xspace}
\newcommand{\yarrp}{Yarrp\xspace}
\newcommand{\seeu}{IPvSeeYou\xspace}
\newcommand{\maxm}{MaxMind\xspace}
\newcommand{\ds}{dataset\xspace}

\begin{document}
 
\author{
 {\rm Erik C. Rye}\\
 University of Maryland
 \and
 {\rm Robert Beverly}\\
 CMAND 
}
  \title{\huge IPvSeeYou: Exploiting Leaked Identifiers in IPv6 for Street-Level
  Geolocation}

\maketitle
\begin{abstract}

We present \emph{\seeu}, a privacy attack that permits a
remote and unprivileged adversary to physically geolocate 
many residential \vsix 
hosts and networks with street-level precision.  The crux of
our method involves: 1) remotely discovering wide area (WAN) hardware MAC addresses from 
home routers; 2) correlating these MAC addresses with
their WiFi BSSID counterparts of known location; and 3) extending 
coverage by associating devices
connected to a common penultimate provider router.

We first obtain a large corpus of MACs embedded in \vsix addresses
via high-speed network probing.
These MAC addresses are effectively leaked up the 
protocol stack and 
largely represent WAN interfaces of residential 
routers, many of which are all-in-one
devices that also provide WiFi.  
We develop a technique to statistically infer the
mapping between a router's WAN and WiFi MAC addresses 
across manufacturers and devices, and mount
a large-scale data fusion attack that correlates WAN MACs with WiFi BSSIDs
available in wardriving (geolocation) databases.  Using these correlations, we
geolocate the \vsix prefixes of 
$>$12M
routers in the wild across 146 countries and territories.  Selected
validation confirms a median geolocation error of
39 meters. We then exploit technology and deployment constraints 
to extend the attack to a larger set of \vsix residential routers
by 
clustering and associating devices with a common penultimate provider router.  
While we responsibly disclosed our results to several manufacturers 
and providers,
the ossified ecosystem of deployed residential cable and DSL routers
suggests that our attack will remain a privacy threat into the foreseeable
future.
\end{abstract}

\IEEEpeerreviewmaketitle

\section{Introduction}
\label{sec:intro}

\ac{MAC} addresses are designed to be globally
unique layer-2 network interface hardware identifiers.  Most modern
network interfaces, including Ethernet, \wifi, and \bt, utilize
48-bit IEEE MAC addresses~\cite{ieee802}.  For several well-known reasons -- notably
manufacturer fingerprinting~\cite{acsac16furious} and the
ability to track devices by an identifier that remains static across network 
changes~\cite{vanhoef2016mac,fenske2021three,cunche2014know} -- MAC addresses are considered
sensitive.  

MAC addresses are typically confined to layer-2, and thus cannot
readily be discovered by a remote attacker who is not attached to
the same subnet.  A historical exception is the use of 
MAC addresses to automatically select the host bits of an \vsix
client, a process known as SLAAC \eui addressing~\cite{rfc2462}.  
Due to the aforementioned vulnerabilities, modern operating systems instead
typically generate IPv6 addresses with random host bits~\cite{rfc4941,rfc7217,rfc8981}.

Prior work in high-speed active \vsix network topology 
techniques~\cite{imc18beholder} has 
helped 
overcome the challenge of finding active hosts and networks amid
the vast \vsix address space~\cite{Gasser:2018:CEU:3278532.3278564}.
Recent research in \vsix periphery
discovery~\cite{rye2020discovering} produced a large corpus
of \cpe devices,
\ie residential home cable and DSL
modems providing \vsix service.  Surprisingly,
more than 60M of these
\cpe deployed in the Internet use legacy \eui addresses, likely because they run
older operating systems and legacy configurations inherent in embedded
devices.  

Beyond this relatively minor privacy weakness, our key insight is
that many of these \cpe devices are System-on-a-Chip (SoC) designs,
\eg~\cite{broadcom3390}, with
multiple network interfaces where \emph{each interface is assigned a MAC
address predictably from a small range.}  For example, an all-in-one
device with a \wan, Local Area Network (LAN), and \wifi interface where the \wan address is
\texttt{AA:BB:CC:DD:EE:01}, 
the LAN address is 
\texttt{AA:BB:CC:DD:EE:02}, 
and the \wifi BSSID (Basic Service Set Identifier; the \ac{AP} wireless MAC address) is
\texttt{AA:BB:CC:DD:EE:03}. While the number of MAC addresses
allocated and offsets can differ widely across devices,
manufacturers, and implementations, we 
develop an inference technique that permits us to predict the 
most likely BSSID given the \wan MAC address.

We can then search for the BSSID in available
public wardriving~\cite{wirelessinfidelity,hurley2004wardriving} databases, \eg \wigle~\cite{wigle}, 
Apple Location Services~\cite{applegeo}, and others~\cite{mylnikov,openwifi}.
The ability to bind a CPE \vsix address %
to its corresponding \wifi BSSID leads to our core
contribution: \emph{street-level geolocation} of the \vsix
network prefixes assigned to these \cpe
(cf.\ Figure~\ref{fig:process}).

Our geolocation inferences are not limited to devices
and implementations using legacy \eui addressing.  Where \eui and
non-\eui devices are both deployed in a provider, we cluster those
devices connected to the same upstream provider router to establish
a feasible location for non-\eui devices.  Thus, a single
\eui device connected to a provider router may potentially compromise the
geolocation privacy of \emph{other} customers that are also connected to that
provider router.  Because \cpe software is rarely upgraded, and the
devices themselves are infrequently replaced, we expect our findings
to remain a threat into the foreseeable future.
Our privacy attack, \emph{\seeu}, makes the following primary contributions:
\begin{itemize}
 \item An algorithm to infer, per-CPE manufacturer and device, the offset between
       its \wan and \wifi interface MAC addresses 
       (\S\ref{sec:methodology}).
 \item Validation of \seeu on a subset of geolocation inferences with a median error of
       39 meters, suggesting our technique is accurate and precise
       (\S\ref{sec:validation}).
 \item Street-level geolocation of 12M \vsix \cpe -- and the \vsix
     customer prefixes they connect -- across 146 countries\footnote{We count
       ISO-3166-1 two-letter country codes throughout and use the term
        ``countries,'' although some are dependent territories~\cite{isocc}.}
       using our offset algorithm and performing data fusion
       with wardriving databases
       (\S\ref{sec:results}).
 \item Extension of the technique to additional \vsix \cpe by clustering
       geolocated devices and associating them with non-\eui devices
       (\S\ref{sec:infra}).
 \item Disclosure to
       several equipment manufacturers and service providers, and
       steps toward
       remediation
       (\S\ref{sec:remediation}).
\end{itemize}

\replaced{
While our attack affects a large subset of deployed \vsix routers, primarily
residential devices,
several conditions must be satisfied in order to successfully geolocate a
router with \seeu: a router must 1) be responsive to active probes; 
2)
use \eui \vsix addresses (\S\ref{sec:euicorpus}); 3) have a predictable
offset between the WAN MAC address and its
BSSID (\S\ref{sec:offset}); and 4) have a BSSID in a geolocation database we
query (\S\ref{sec:wificorpus}). We discuss
these limitations further in~\S\ref{sec:limitations} and
 ethical considerations of our work in~\S\ref{sec:ethics}
}{We discuss limitations of \seeu in~\S\ref{sec:limitations} and
 ethical considerations of our work in~\S\ref{sec:ethics}.}

\section{Background and Related Work}
\label{sec:background}

\subsection{\vsix Addressing}

\begin{figure}[b]
    \centering
    \resizebox{0.7\columnwidth}{!}{\includegraphics{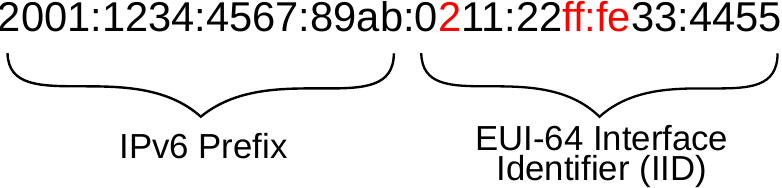}}
    \vspace{-1mm}
    \caption{An \eui \vsix address constructed by embedding the MAC 
    \texttt{00:11:22:33:44:55} in the IID.}
    \label{fig:eui}
    \vspace{-3mm}
\end{figure}

Devices
commonly auto-generate their interface \vsix
addresses through \ac{SLAAC}~\cite{rfc1971,rfc2462,rfc4862,rfc7217,rfc8981}
rather than \replaced{via}{static assignment} assigning
addresses statically or \deleted{via} DHCPv6. 
Early \vsix standards
encouraged the use of \eui \vsix addresses~\cite{rfc3513,rfc4291}, wherein the lower
64-bits of the 128-bit address -- the \ac{IID} -- embed the
interface's \ac{MAC} address.  The embedding (a modified \eui)
first sets the Universal/Local bit, then inserts the bytes \texttt{0xFFFE}
between the third and fourth bytes of the \ac{MAC}. 
Figure~\ref{fig:eui} displays
an example \eui \vsix
address.

Modern devices, particularly end systems, no longer employ \eui \ac{SLAAC}
addressing for several reasons. First, a static, unique \ac{IID} 
allows an adversary to track devices over time and
address space changes. Second, MAC addresses are globally unique, 
with contiguous blocks of $2^{24}$ bits (known as \acp{OUI})
assigned to manufacturers.
Not only does
embedding the \ac{MAC} address in the \ac{IID} expose the device
manufacturer, work has shown that it is possible to infer the specific 
device model~\cite{acsac16furious}.
Instead, modern operating systems typically form \vsix addresses using
randomly generated \acp{IID}~\cite{rfc4941,rfc7217,rfc8981}.

Despite known security and privacy issues inherent in \eui
addressing, and the 
introduction of SLAAC Privacy Extensions (PE)~\cite{rfc3041} over 20 years ago,  
previous
studies~\cite{imc18beholder,Gasser:2018:CEU:3278532.3278564,rye2020discovering}
discovered millions of \ac{CPE} devices using \eui \ac{SLAAC}.
Our work focuses on these devices, 
which primarily include residential home cable and DSL
routers.

\subsection{IP Geolocation}

IP addresses are logical network identifiers; while addresses 
and hostnames may
identify a network or operator and hint at the location, the
associated device may physically be anywhere.  Further, the
device may not wish to reveal its location,
or may be unable to geolocate itself.  As a result, a rich body of
work has developed IP geolocation techniques that allow a
third-party to map arbitrary IP addresses to physical locations.
Multiple IP geolocation services, \eg
\cite{edgescape,maxmind,hexasoft}, exist to support applications
such as advertising,
content and language customization, content geo-fencing, law and
policy enforcement, anti-fraud, authentication, and 
forensics~\cite{katz_2006,huffaker2011,wang_2011}.

Well-known methods for third-party IP geolocation include: 1) registry databases,
\eg whois and the DNS~\cite{10.1145/383059.383073}; 2)
constraint-based techniques that leverage speed-of-light propagation delay
to triangulate an address~\cite{gueye2006constraint,10.1145/2398776.2398790}; 
3)
network topology~\cite{katz_2006,10.1145/2381056.2381058}; and 4) privileged
feeds~\cite{rfc8805}.

While these geolocation services impinge on the privacy of the devices
and users, they generally provide 
course-grained location, \eg city.  
Several studies have found significant inaccuracies in 
techniques and databases as compared to ground truth.
For instance, Poese \etal found 50-90\% of
ground truth locations present in commercial databases had
greater than 50 kilometers of
error~\cite{poese2011ip}; more recently Komosn{\`y} \etal studied
eight commercial geolocation databases and found mean errors ranging
from 50-657 kilometers~\cite{komosny2017location}.

In contrast to these prior techniques, our approach seeks to
geolocate \vsix addresses
at a street-level granularity.
While Wang \etal similarly
sought street-level geolocation, albeit for \vfour, their technique requires 
geolocation targets to reside in a high-density location and
nearby permissive passive landmarks~\cite{wang_2011}.
Finding acceptable landmarks, even in dense locations, is a
significant hurdle in the modern age of shared hosting services.  In
this paper, we show that not only can \seeu provide \deleted{highly}
accurate street-level geolocation, it is effective on a large number
of \vsix prefixes.

\subsection{Related Work}

\deleted{In~\cite{wright2015hacking},}
Wright and Cache observed that the \wifi and \bt
\ac{MAC} addresses of mobile devices are often sequential,
allowing passive adversaries to correlate
these identifiers~\cite{wright2015hacking}. Our work also leverages the
idea of predictable MAC assignment 
across different link-layer technologies, but 
does not require physical proximity to the
target and focuses on \cpe rather than mobile devices.

The
extraordinarily large address space of \vsix removes the need for
\ac{NAT}.
Whereas \ac{NAT} is ubiquitous in residential \vfour 
networks, \vsix restores an end-to-end connectivity model whereby
the \ac{CPE} device is a routed hop.
This
requires a novel approach to \ac{CPE} discovery in \vsix.
Our ``edgy'' algorithm is
specifically aimed at discovering the \vsix network periphery, \ie the \ac{CPE} that connect customer edge networks to the
\vsix Internet~\cite{rye2020discovering}.  With edgy, we previously
discovered 5M
unique MAC addresses in 16M \eui \vsix addresses, but did not attempt to
correlate these MAC addresses with wireless identifiers or geolocate
them, as we do in this work.

Recent work has sought to understand \vsix addressing.
Fiebig
\etal~\cite{pam2017-nxdomain} and Borgolte \etal~\cite{BorgolteSP2018} 
used DNS
response semantics to discover active addresses within reverse zones. Murdock
\etal~\cite{Murdock:2017:TGI:3131365.3131405} 
generated target addresses  and test for liveness using active
measurements.
Li and Freeman~\cite{li2020towards} examine user-level \vsix behavior 
and address dynamics from the
vantage of a large online social network, 
and how to best implement effective \vsix filtering. 

Prior work has exploited \vsix to mount tracking campaigns.
Berger \etal reversed the keyed hash function used to generate
\vsix flow labels, thereby permitting device tracking~\cite{berger2020flaw} 
even when the device uses random addresses; this vulnerability has
since been mitigated by common operating systems.
More recently, Rye \etal leveraged CPE \vsix \eui addresses to
track connected devices 
across prefix and \ac{IID} changes~\cite{rye2021follow}. 
Our work similarly exploits \vsix \eui addresses to attack user
privacy, but via
precision geolocation for a subset of users and devices.
While limited prior work has examined \vsix geolocation~\cite{tran2014},
to the best of our knowledge \seeu is one
of the first techniques to exploit specific properties of \vsix
for geolocation.

\section{Methodology}
\label{sec:methodology}

\begin{figure}[t]
    \centering
    \resizebox{0.7\columnwidth}{!}{\includegraphics{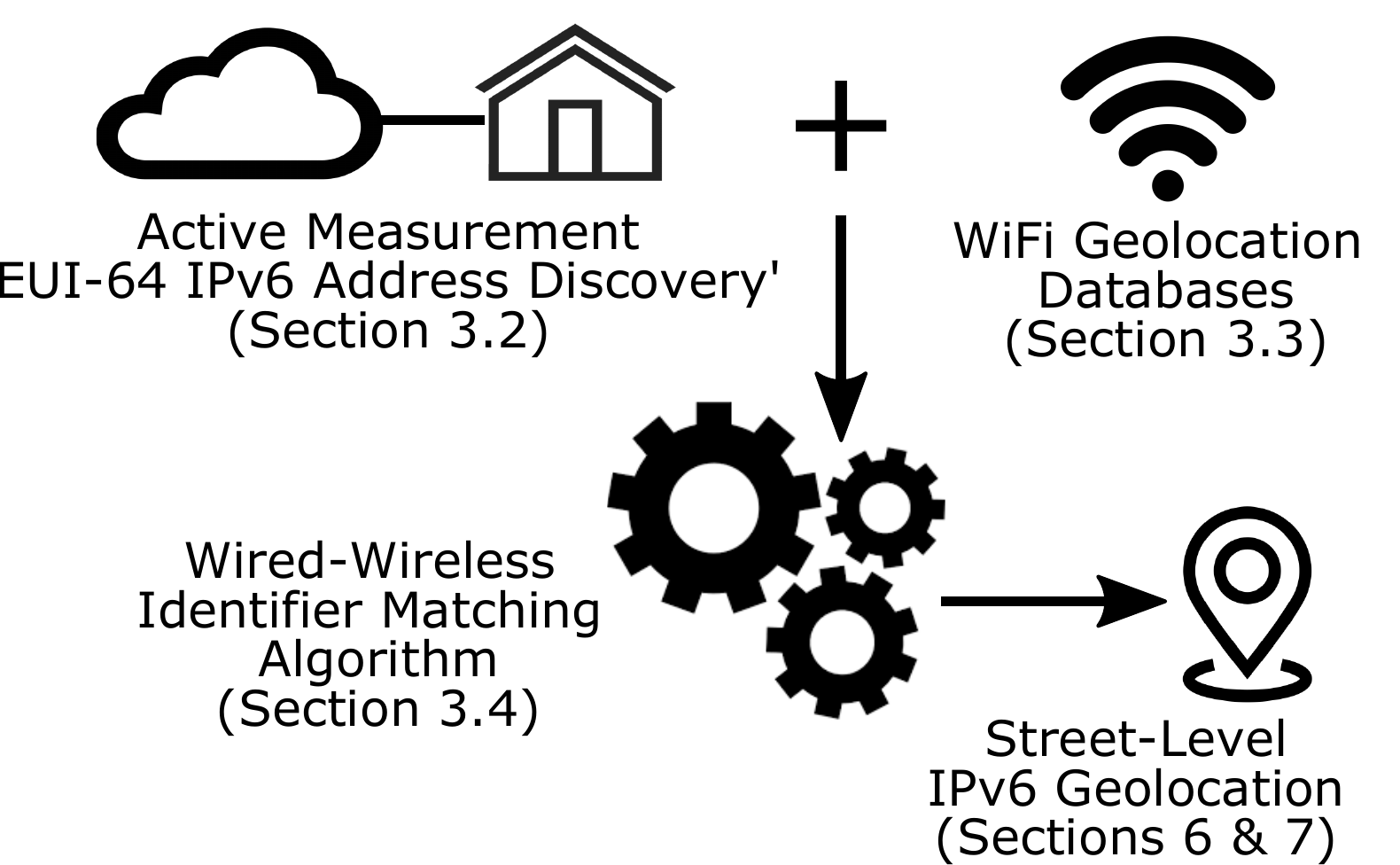}}
    \vspace{-2mm}
    \caption{\seeu: to geolocate \vsix \cpe, we fuse %
    MAC addresses %
    from \eui \vsix addresses (\S\ref{sec:euicorpus}) 
    with \wifi BSSIDs from
    geolocation databases 
    (\S\ref{sec:wificorpus}). Our matching algorithm (\S\ref{sec:matching})
    produces an inferred per-OUI offset. %
    }
    \label{fig:process}
    \vspace{-2mm}
\end{figure}

Our starting point exploits \cpe implementations with two key facets:
Internet-facing WAN interfaces that use \eui addresses, 
and predictable
MAC address assignment across the device's wired and wireless
interfaces.
\eui addressing allows an adversary to remotely obtain MAC 
addresses from vulnerable devices by eliciting responses from network probes
(\eg traceroute, \yarrp~\cite{imc16yarrp}, zmap6~\cite{durumeric2013zmap, zmap6}). 
Predictable MAC address assignment allows
an adversary to map wired MAC addresses obtained from active network
probing to wireless BSSIDs from \wifi geolocation databases. %
While this section
focuses on exploiting \cpe that use \eui addresses, we extend the technique's
potential coverage to %
\cpe that do not use \eui SLAAC in \S\ref{sec:infra}.%

Figure~\ref{fig:process} outlines our methodology; in \S\ref{sec:euicorpus} we
discuss our \eui \vsix corpus, \S\ref{sec:wificorpus} describes our BSSID
geolocation data, and \S\ref{sec:offset} gives our algorithm for
linking \eui \vsix-derived MAC addresses with BSSIDs.
First however, we provide an example of
\seeu to build intuition.

\subsection{Example}

\begin{figure}[t]
    \centering
    \resizebox{0.9\columnwidth}{!}{\includegraphics{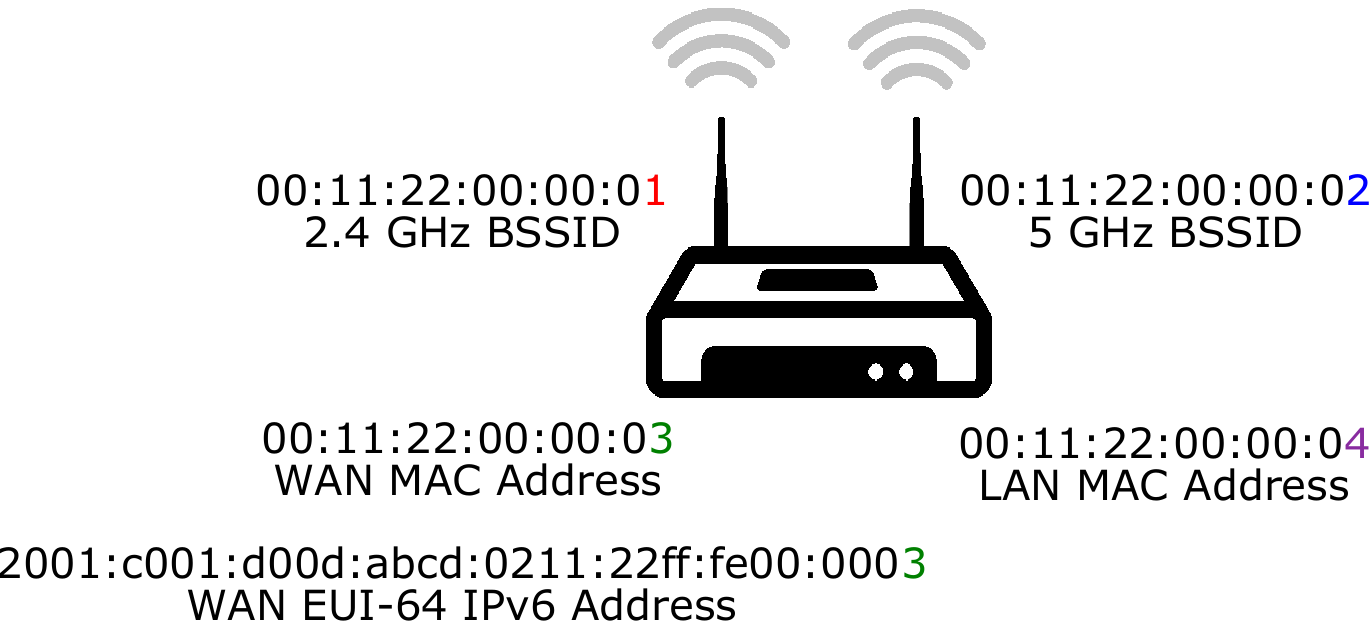}}
    \caption{A vulnerable \ac{CPE} router.
    The four interfaces (2.4 and 5 GHz 802.11 BSSIDs, LAN and WAN)
    are assigned sequential MAC addresses. The
    \eui SLAAC WAN \vsix address %
    is
    discovered via active network scans, while the BSSIDs
    are found in %
    \wifi geolocation databases.}
    \label{fig:router}
\end{figure}

Figure~\ref{fig:router} depicts a router that is vulnerable to
the \seeu geolocation technique. In this simple example, each router interface 
is addressed sequentially from
the same \texttt{00:11:22} OUI; furthermore, it generates its WAN \vsix address
using the modified \eui derived from the WAN interface MAC address. This allows its WAN
MAC address to be discovered by active network measurements that elicit a
response from the CPE. 
Its \wifi BSSIDs
are contained in crowdsourced databases, such as \wigle~\cite{wigle}, that give a
precise geolocation for the BSSID.
The offset distance between the WAN MAC
address and BSSIDs (in Figure~\ref{fig:router} the offsets are -1 and -2) typically remains fixed
throughout an OUI, or at a minimum, for device models within an OUI.
This allows an attacker who has discovered CPE MAC addresses from active network
scans to \emph{predict} the device's BSSID(s) and look them up in \wifi geolocation
data, or \emph{fuse} previously-obtained data sources together.

\subsection{\vsix \eui Corpus}
\label{sec:euicorpus}

Our previous work on \vsix ``periphery discovery''~\cite{rye2020discovering} 
employs an
iterative, targeted scanning algorithm to find CPE routers
using Yarrp~\cite{imc16yarrp}.  We refine this original technique to
also use non-TTL limited ICMP6 echo request probes and performed
additional Internet-wide scanning campaigns from July 2020 through July 2021.

The resulting \vsix periphery discovery corpus 
includes a large number of
\eui \vsix 
responses.
\eui \vsix addresses are readily identifiable and easily reversible -- the MAC addresses are
decoded from \eui response addresses by removing the fourth and fifth bytes
(\texttt{0xFFFE}) of the \ac{IID}, then inverting the U/L bit.
The corpus contains 
approximately 347M
\eui \vsix addresses with
nearly 61M
unique \ac{MAC} addresses. Note that \ac{MAC} addresses can appear in
multiple \eui \vsix addresses due to ephemeral prefix leases provided through
temporary-mode DHCP~\cite{rye2021follow} and devices moving to new
networks. Smaller numbers of repeated MAC addresses occur due to
address reuse.%

Of the 61M total MAC addresses
derived from the \eui \vsix \ds, approximately 0.2\% 
(126,730) were observed in
\eui addresses in multiple \acp{AS}. 
While this dispersion may be attributable to customers and
devices changing service providers, we exclude them from
analysis to eliminate potential occurrences of non-unique
MACs.
The Appendix provides a 
detailed analysis of the OUI and AS distribution of
the WAN corpus.
\begin{figure*}[t]
 \centering
 \resizebox{0.9\width}{!}{\includegraphics{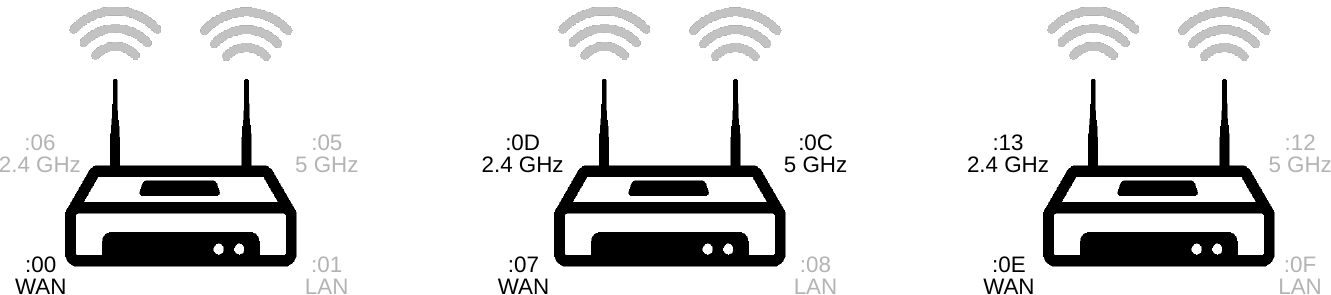}}
 \caption{Example where three WAN MACs and three \wifi BSSIDs are observed
          (first five bytes omitted; observed MACs shown in bold).
          We first infer the number
          of MAC addresses allocated per device based on all data
          observed within the OUI, then compute the most
          likely WAN-to-BSSID \emph{offset}.  Missing data and multiple in-block
          BSSIDs (\eg guest \wifi and different frequencies)
          complicate the inference.  
          Each device is
          allocated seven MAC
          addresses, and the offset is +6.
          Note that a na\"ive closest matching 
          (\eg between \texttt{0x0D} and \texttt{0x0E}) is
          incorrect.
}
 \label{fig:fritz}
\end{figure*}

\subsection{\wifi Geolocation Data}
\label{sec:wificorpus}

While our methodology is agnostic to the source of geolocation data,
this study uses five sources, including three
open-source databases~\cite{mylnikov, openwifi, openbmap}, the
\wigle API~\cite{wigle}, and Apple's \wifi geolocation
service~\cite{applegeo}. 

The three databases contain 20M (Alexander
Mylnikov)~\cite{mylnikov}, 15M (OpenBMap)~\cite{openbmap}, and 29M
(OpenWifi.su)~\cite{openwifi} BSSIDs respectively, along with associated geolocations.
Because these databases rely on
crowdsourcing, they are biased toward the locations of
contributors. 

In addition to the three databases,
we issued wildcard queries for the OUIs in our corpus via the \wigle
API.
As the standard API rate-limits were
prohibitive, we coordinated with the \wigle administrators to
increase our daily query limit  to
obtain
1,367,700 geolocated BSSIDs. 
As with the other databases, \wigle's coverage is dependent on 
the location of the crowdsourcing contributors.

Finally, we also obtain BSSID geolocation data using Apple's \wifi geolocation
service~\cite{applegeo}. Apple provides this API for its products to geolocate
themselves as part of its Location Services suite of tools; the API accepts
an 802.11 BSSID as a search parameter. If Apple has geolocation
data for the BSSID, it returns these data, optionally with additional location information
for \acp{AP} in close proximity. The purpose of
returning the additional geolocation information is presumably to short-circuit
API requests from the same client as it encounters these additional nearby
\ac{AP} BSSIDs.

We used the Apple geolocation API as an oracle to validate the existence
of BSSIDs we suspect are related to our \eui MAC addresses. We queried the
location service for BSSIDs at offsets increasing from 0 from our wired MACs
from~\S\ref{sec:euicorpus}. When we guessed a valid BSSID, the geolocation
service returns not only the coordinates of the guessed BSSID, but additionally
up to 400 nearby BSSIDs and their geolocations~\cite{isniff}. We stopped
querying for BSSIDs within an OUI if no offset value between successive WAN MAC
addresses produced a valid BSSID. This results in our largest \wifi geolocation
\ds, with 444,860,460 unique BSSIDs. 

In total, our geolocation data contains 450,018,123 distinct BSSIDs in
238 countries and territories. 
We use the IEEE OUI database
to map OUIs to manufacturers~\cite{oui}.
Table~\ref{tab:geostats} in the Appendix summarizes macro-level characteristics
of the geolocation data.  Given the potentially sensitive nature of the
data we collect and aggregate, we sought and followed guidance from
our IRB; see~\S\ref{sec:ethics} for details.

\subsection{Inferring WAN-to-BSSID MAC offsets}
\label{sec:offset}

Key to our method is correlating addresses between network interfaces
on a CPE device.  
Given a WAN \vsix address with an embedded MAC address, we wish to determine the MAC address of a
corresponding \wifi interface on the CPE.  
In the trivial case,
the \wifi BSSID MAC is exactly one greater than the WAN MAC address.
However, the assignment of
MAC addresses is vendor and device dependent.  
For
example, Figure~\ref{fig:router} shows a CPE with four interfaces: 
LAN, WAN, and two different
\wifi radio frequencies.  In this example, the two BSSID MAC address values are
one and two less than the WAN MAC address.  In practice, more complex allocations
exist and there is a wide variety of deployed implementations.

\subsubsection{Challenges}

To enable our data fusion, we require a mapping of the offsets between
interface MAC addresses on a per-OUI basis.  Unfortunately, vendors do
not publish their MAC address assignment policy, and even a single
vendor frequently uses different strategies for different devices.

Thus, given the huge variety of vendors and deployed CPE devices in
the Internet, we develop an algorithm to statistically infer the MAC address
offsets.  We utilize the large number of WAN and \wifi MAC addresses
in our corpora (summarized in the Appendix)
to capture this diversity and build a database of
offsets for different devices.

A na\"ive approach to building the mapping is simply associating a
WAN MAC with the numerically closest BSSID.  Such an
approach can fail simply due to missing data points; 
for instance, if only the device's WAN address is present in our
data, or only the BSSID is present.  In these cases a false association
can be made where the two linked addresses belong to different 
devices.  These errors can be mitigated in part by inferring the
number of addresses allocated per device and preventing associations
between two addresses that differ by more than the size of the 
allocation.

More subtle errors can, however, occur. 
Figure~\ref{fig:fritz} illustrates how a simplistic algorithm fails
for a particular CPE we purchased and for which we have ground truth
(a CPE from the vendor AVM, which is prevalent in our data).
This device uses a block of seven contiguous MAC addresses for its
various interfaces\footnote{While a block size of seven is immediately
conspicuous for being odd, prime, and not on a nybble boundary, both
our ground-truth and inference algorithm reveal that this is the true
manufacturer allocation policy.}.  The lowest MAC address is given to the WAN
address while the highest is given to the 2.4 GHz WiFi interface.
Thus, the true offset is +6.  Because of this allocation, the nearest
match association can result in the WAN address of one device being
associated with the BSSID of a different device; for example 
MACs ending in \texttt{0x0D} and \texttt{0x0E} in Figure~\ref{fig:fritz}.  Also in the
data are MACs corresponding to different \wifi radio frequencies.  For
instance, this model of device also has a 5 GHz WiFi interface, but at
an offset of +5.  

Further, our data may include a single MAC address for a device, \ie
either just the WAN or just the BSSID.  Missing data is common, and
can occur simply because %
wardrivers never encounter a device,
a network blocks our probes, or there is 
other filtering.  For example, Figure~\ref{fig:fritz} shows
one device where our data includes only the WAN MAC (at
position \texttt{0x00}).  In these cases, the nearest matching BSSID may be
a multiple of the true offset, for instance +12.  Missing data,
multiple in-block BSSIDs, and very sparse or dense OUIs therefore
complicate the inferences.

\subsubsection{Algorithm}
\label{sec:alg}

Our algorithm infers the most likely offset between the 
WAN MAC address and BSSID for a given OUI.  
First, we determine the OUI's mostly likely
allocation size division, \ie how many MAC addresses are allocated
per device.  We sort all of the BSSIDs in the OUI to build a
distribution of intra-MAC distances.  Thus, for $n$ input BSSIDs, we
compute $n-1$ distances between these points.  We find the most
frequent distance and then determine how many of the samples in the
distribution correspond to a multiple of this distance by computing
the greatest common divisor (gcd).  If the fraction of distances that
are multiples of this distance are high, then we correspondingly have
high confidence that the inferred allocation size is correct. 

Given the inferred allocation size, the algorithm next
iterates through each \eui MAC address in ascending sorted order for every 
OUI with at least 100
WAN MAC and 100 BSSID instances.
Because the matching \wifi MAC address
may be at either a positive or negative offset, the algorithm finds
both the closest corresponding BSSID less than, and greater than, the
\eui-derived MAC, subject to the constraints that these must be within a window
determined by the inferred allocation size in the previous phase.
Finally, the algorithm infers the offset for this device to be
the most common offset among all the matches.  

During execution of this algorithm, both correct and false
associations will be made, 
for
instance the false association to the -1 offset BSSID versus the +6
offset BSSID in the example of Figure~\ref{fig:fritz}.  However, the
intuition is that, in aggregate,
it will be more common for a single device to be
present in the data with \emph{both} its addresses than for two different
devices with adjacent addresses.  While exceptions can exist,
especially for OUI with a large number of devices present in our data, in
practice, statistically choosing the offset produces the correct
inference for our ground truth devices.  We again compute the fraction
of devices for the OUI that conform to our inferred offset such that
we can have an associated confidence measure.

\subsubsection{Correlating \vsix \eui MACs and BSSIDs}
\label{sec:matching}

Given a MAC address embedded in an \eui \vsix address, the final step
in our technique is to utilize our offset database\deleted{, as described
in the previous subsection (\S\ref{sec:alg}),} to look up the offset to the BSSID 
given the OUI in question.  We then lookup the corresponding inferred 
BSSID in the 
wardriving databases to make our final geolocation inference.  Note
that if the OUI is not contained in our offset database, we 
cannot make a geolocation determination.

\subsubsection{MAC-to-BSSID offset confidence}

\begin{figure}[t]
    \centering
    \resizebox{0.8\columnwidth}{0.5\columnwidth}{\includegraphics{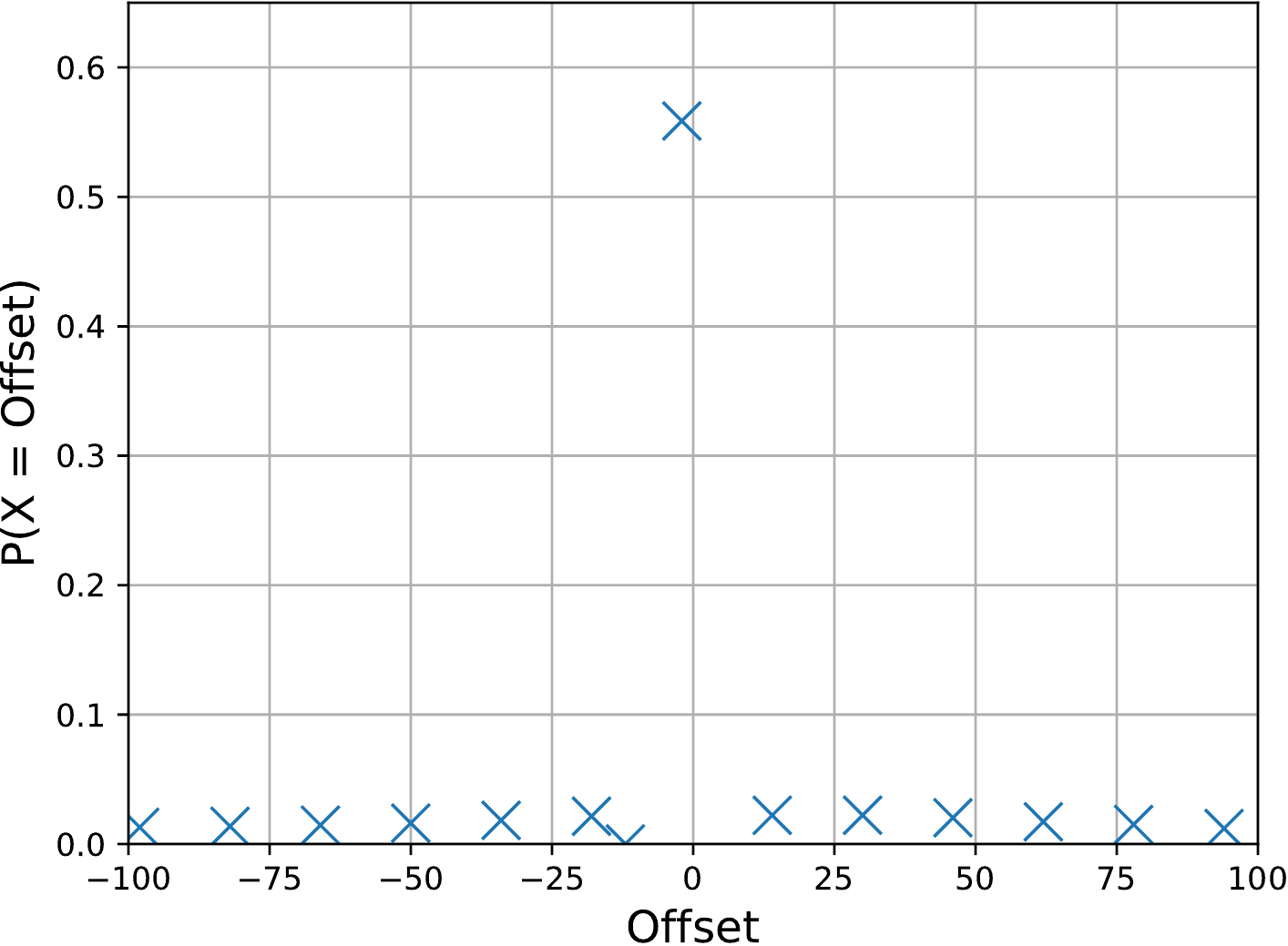}}
    \caption{Probability mass of offsets between observed
    Arris addresses in the \texttt{00:1D:D1} OUI.  The peak offset is -2 with other points
    at multiples of 16 away, \ie harmonics of the -2 offset.  99.9\%
    of the probability mass supports the inferred -2 offset, with
    15,165 total offset inferences.}
    \label{fig:arrisoffset}
    \vspace{-3mm}
\end{figure}

As Figure~\ref{fig:fritz} illustrates, offset inference is 
complicated by manufacturer differences and available 
observations. Assuming a
relatively dense number of observations of WAN addresses and
BSSIDs, our algorithm intuitively accumulates the most ``matches''
at the correct offset and lesser counts at regular intervals (+/-
the number of MAC addresses allocated to individual devices). In
addition to 18 ground-truth devices we purchased
(\S\ref{sec:validation:used}), we employ
statistical measures to infer and validate offsets for each OUI.
For instance, Figure~\ref{fig:arrisoffset} displays a
PMF of the offset value for an Arris OUI.  %
The peak offset is
-2, while smaller impulses occur at intervals 16 addresses away.
In this particular case, Arris allocates a span of 16 MACs to
each device. Thus, the other points are a \emph{harmonic} of the -2
offset (where the distance is to a different device).  Indeed,
99.9\% of the probability mass in this plot supports the -2
inference.  

\section{Limitations}
\label{sec:limitations}

While our methodology allows us to make geolocation inferences
for a large number of \vsix networks -- we use \seeu to 
geolocate more than 12M routers in the wild -- it is limited to
specific CPE behaviors and deployments.  %

First, \seeu relies on CPE that use \eui WAN \vsix addresses
and are responsive to our active probing.
Second, \seeu is only effective against all-in-one CPE that 
include a built-in \wifi base station.  In contrast, some home
networks contain a standalone cable modem with an Internet-facing
\eui \vsix address that is then connected by Ethernet to an \ac{AP}.
In this case, it is impossible for us to predict the BSSID from
the WAN MAC address, as they are two entirely distinct devices.

While many all-in-one CPE use a single integrated SoC,
our associations are limited to devices where all interfaces
are allocated MACs from the same OUI.  
If the device's BSSID resides in a
different OUI, our methodology will not find any potential offsets,
and thus will be ignored.  

Further complicating the offset inference are instances where
addresses from a single OUI are divided among multiple device
models, a phenomenon observed by Martin~\cite{acsac16furious}. 
If multiple
devices with more than one distinct WAN MAC address to BSSID offset
value exist within a single OUI, our algorithm does not capture this
nuance, instead choosing the predominant offset value as learned from
the data. However, our confidence measure identifies these
heterogeneous OUIs as problematic.  While we ignore OUI with
low-confidence offsets, a more sophisticated algorithm could
additionally infer these granular OUI allocations 
in the future.

Finally, the underlying BSSID geolocation data we use to geolocate a \vsix
address may itself be incorrect or outdated. 
Additionally, devices geolocated in the past may have since moved,
introducing inaccuracy in our geolocations.

While these constraints limit the attack's scope, we note that:
1) all-in-one CPE with predictable offsets are common;
2) the validation we perform, while limited, confirms both the
technique's viability and accuracy;
and 3) we extend the technique's coverage 
in \S\ref{sec:infra} to associate CPE which do not meet the above
constraints with \emph{other} CPE that do.  

\section{Ethical Considerations}
\label{sec:ethics}

Fundamental to our work are MAC addresses, which uniquely identify
network interfaces, and, hence, devices.  While not a user identifier
per se, MAC addresses can be leveraged to track users or combined with
other meta-data -- as we explicitly show.  As such, we submitted our
research plan and protocols \replaced{to}{with} our institution's IRB, who cleared the 
study.  Our IRB noted that we 
have no way to associate any of
our data with individuals, and that there was the potential for
\deleted{large} overall societal benefits
from the research by improving privacy for millions of residential
Internet users.

To minimize risk, we treat MAC addresses and  
any correlated geolocations as private data.  We only publish,
share, or release aggregated analyses on the data, and ensure 
that the raw data at rest remains encrypted.  

While the results of our research could be misused, we aim to
ultimately improve privacy protections by highlighting this
vulnerability.  In addition, we have responsibly disclosed the privacy
weaknesses of exposing MAC addresses in \vsix to network
equipment vendors and a large residential service provider.  At least one vendor is in the process of issuing a patch to update
their equipment's behavior, and the residential service provider is
currently deploying measures to mitigate our attack.  In light of these factors, we believe, as
does our IRB, that the
beneficence of our work significantly outweighs any potential harm or
risk it may present.
\section{Validation}
\label{sec:validation}

We employed a multi-pronged strategy for validation, including
crowd-sourcing measurements, 
purchasing selected CPE devices, 
and
collaborating with a large residential ISP. %

\subsection{Crowd-sourced Measurements}
\label{sec:euivols}

\replaced{Our first validation experiment}{We} enlisted the help of
volunteers\deleted{ to validate
our methodology}. We designed a custom web page that first tests for
\vsix connectivity \added{and, i}f the client has operational \vsix,
\deleted{the web site}
logs \added{the} client's address and requests the client's location 
via the HTML5 geolocation API.  If the user consents,
we obtain tuples of \vsix address and precise
geolocation.
From
the client address, we obtained their CPE router's \vsix
address via active probing (edgy) and employed the \seeu inference procedure.  
We publicized our measurement site via \replaced{the SIGCOMM Slack channel}{social media}.
Of the 50 participants with residential \vsix
service from 31 \acp{AS} in Europe, North America, and Asia that participated in
our user study, the majority (84\%) did not have CPE using \eui \vsix
addresses. Of the eight with \eui \vsix addresses, \seeu
successfully geolocated five.

The true geolocation for four of these five CPE
agreed with \seeu's inference 
within 50 meters. By contrast, \maxm's GeoLite2 geolocation database
geolocated these addresses
to between 500 meters and 421 kilometers from the true location, with a mean
error of 106 kilometers and median error of
1.34 kilometers.
The \seeu geolocation of the fifth device was 0.68 kilometers from the
HTML5
geolocation, while the \maxm location was over
300 kilometers from both the HTML5 and \seeu geolocations.

For the three devices for which we did not obtain an
\seeu geolocation, one was due to an inability to determine the
MAC address offset.
This occurred because we lacked a sufficient number of
observations of WAN MACs and BSSIDs to make an offset inference. %
\seeu was able to determine the offset for the remaining two CPE,
but the
inferred BSSIDs were not found in our wardriving data. This validation experiment
demonstrates 
\added{that \seeu can provide highly precise geolocations for some
devices,}
\deleted{the accuracy with which \seeu geolocates devices that use both
\eui \vsix addressing and sequential MAC addresses. It} and highlights
some real-world challenges, such as CPE without \eui
addresses and address offset inference failures. 
\added{Due to the small sample size of the crowd-sourced measurements,
we ultimately collaborated with a large North American ISP for more
representative validation, as detailed next.}

\subsection{Provider Validation}
\label{sec:providerval}

We coordinated with a large United States-based residential and
business ISP that offers \vsix to obtain
validation.  From our complete \ds,
we randomly sampled 1,350 responsive CPE addresses from the
provider's address space.  We used a bulk reverse geocoding
service~\cite{latlong-net} to map our inferred geo-coordinates to a
US ZIP code and supplied the $<$IPv6 CPE address, customer
subnet, ZIP code$>$ tuples to the provider.  Of the full set, the
ISP was able to provide validation for 486 CPE (36\%) due to
missing records and address churn.  Of this subset, 80\% of our ZIP
code geolocation inferences agreed with the provider's ground-truth ZIP codes.

Due to stringent customer privacy regulations, the provider could
unfortunately not provide further detail about individual errors or
error bounds.  We find this 80\% agreement encouraging given: 1)
ZIP code geo-granularity is widely accepted as useful for marketing
and demographic correlations (there are approximately 44,000 ZIP codes
in the United States); 2) natural address churn and reuse during the
delay between providing our inferences and receiving validation; and
3) the potential that we inferred a ZIP code adjacent to the correct
ZIP.

\subsection{Ground-truth Hardware}
\label{sec:validation:used}

\replaced{Finally, as a third source of complementary validation, we}
{We} purchased 18 used hardware devices from \acp{OUI} prevalent in our
corpus.  Not only did these ground-truth CPE inform our
offset algorithm, they exposed real-world pathologies and limitations
of \seeu.

Some of the \acp{OUI} of MAC addresses derived from our active scan \ds
have no corresponding BSSIDs in our wireless data; this means that
entire OUIs are ``matchless.''
This behavior is primarily due to manufacturers allocating MAC addresses to the 
wired and wireless
interfaces of a \ac{CPE} from different OUIs. For example, we procured several
Technicolor~\cite{technicolor} CPE routers distributed by
Comcast~\cite{comcast}, for its Xfinity Internet service.
Five devices addressed their wired and wireless interfaces from different OUIs
(\eg \texttt{FC:91:14} and \texttt{78:F2:9E}, respectively.)
Despite a heterogeneous mix of
\acp{OUI} on a single CPE, some patterns still exist. For instance, as the lower
24 bits of the wired interface MACs increase, so too do the lower 24 bits of the
BSSIDs in the \texttt{78:F2:9E} OUI, albeit non-uniformly. 

Among the 18 ground-truth CPE we procured, our offset algorithm
correctly returned no inference for all ten devices with different OUIs for
WAN MAC and BSSID allocations. Among the eight for which our algorithm did return an
offset, five were correctly predicted and three were incorrectly predicted. Further
investigation revealed that the three incorrectly-predicted offset exemplars have no
close MAC observations in our active scan data, potentially indicating the
purchased devices do not use EUI-64 \vsix addresses and thus do not appear in
our corpus.

\begin{figure*}[t]
    \begin{subfigure}[c]{.48\textwidth}
    \centering
    \includegraphics[width=\columnwidth]{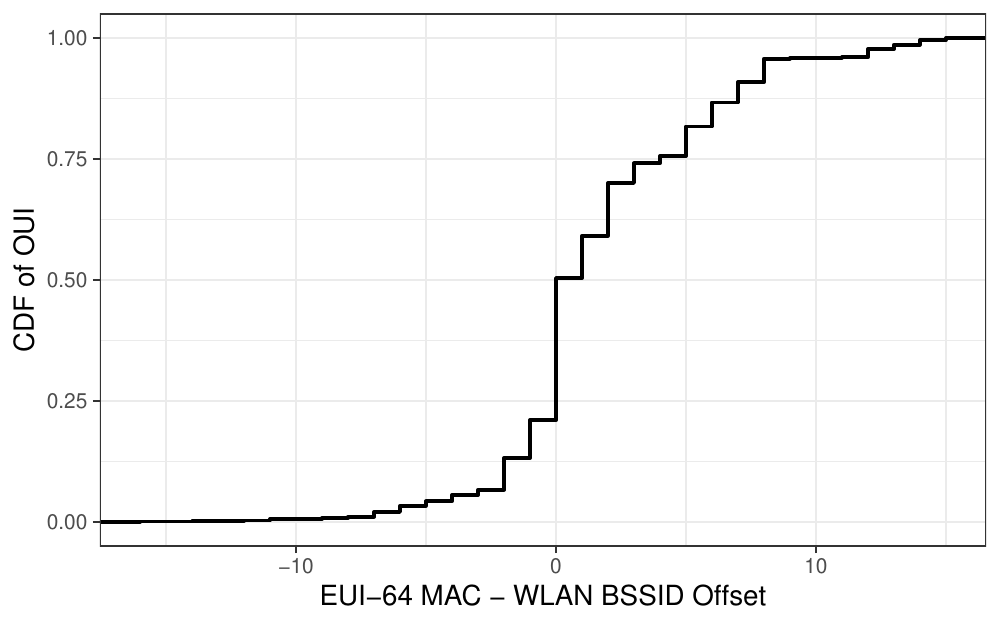}
    \caption{CDF of 1,008 analyzed OUIs' inferred %
      MAC-to-BSSID offsets.
    All inferred offsets fall within the range of -16 to 15;
      0 (the WAN MAC and BSSID are the same) %
      is the most common offset.}
    \label{fig:ouioffsetcdf}
  \end{subfigure}%
    \hspace{1em}
    \begin{subfigure}[c]{.48\textwidth}
    \centering
      \includegraphics[width=\columnwidth]{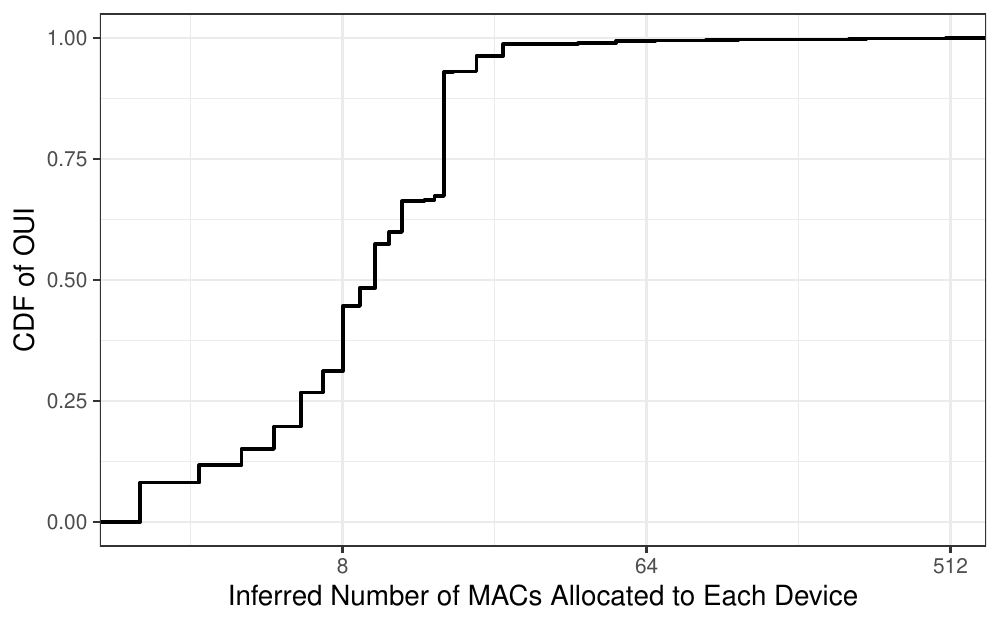}
  \caption{The inferred number of MAC addresses allocated to each device as a
      CDF of the analyzed OUIs. Note that the vast majority of OUIs ($\sim$90\%) have
      inferred allocation sizes of 16 or fewer.}
    \label{fig:alloccdf}
  \end{subfigure}%
  \caption{Inferred wired MAC to wireless BSSID offset and MAC address
    allocation size CDFs.}
\label{fig:allocationandinference}
    \vspace{-3mm}
\end{figure*}

\section{Results}
\label{sec:results}

We next present results of \seeu on our data, first by characterizing the
inferred CPE WAN-to-BSSID offsets and then in terms of the
geolocations it produces.

\subsection{Offsets}

Using our methodology (\S\ref{sec:alg}) we compute the
inferred \wtb offset value for each OUI with more than 100 data points
in the \eui WAN data,
\ie only those OUI where we
have enough data to make a meaningful inference.  We then filter OUI
by the fraction of data points that are consistent with the inferred
offset or a harmonic of the inferred offset.  We apply a liberal
filter that excludes OUI where fewer than 5\% of the data points are
offset consistent.

We discuss some scenarios that may cause an OUI with many \eui-derived
MAC addresses to have few or no matches in \S\ref{sec:limitations}. 
Our rationale for only filtering the lowest-confidence offset OUI is
to accommodate instances where the OUI contains multiple device models
that implement different offsets, as well as to handle instances where
there are a large number of observations in one of the two media
(wired MACs or wireless BSSIDs) but few in the other.  

After filtering, 1,008 unique OUIs remain, or
\replaced{3\% of the 32,345 registered OUIs (see
Appendix for additional details.)}{only 0.2\% of the 463,188 \eui-derived OUIs.}
These 1,008 OUIs cover 31,720,611 distinct \eui-derived (WAN) MAC
addresses in the corpus, approximately 52\% of the total discovered 
from probing (\S\ref{sec:euicorpus}). Of these $\sim$31M \eui MAC addresses, 12,125,839
have a predicted BSSID found in our geolocation database (a ``match'')
at the inferred offset value derived from
our matching algorithm.
This represents $\sim$38\%
of the \eui-derived MAC addresses from the 1,008 filtered OUI and $\sim$20\% of
the entire \eui-derived MAC address corpus. 

Figure~\ref{fig:ouioffsetcdf} depicts the cumulative fraction of
inferred offset value across OUIs
we analyze. 
The inferred offsets
range between -16 and 15, with a
statistical mode of zero.
The range of inferred offset values is unsurprising; device manufacturers assigning MAC
addresses sequentially to interfaces will naturally produce small offset distances between
any two, and the range suggests that typically single-device MAC addresses do
not stray more than a nybble away from each other. More surprising is the
slightly more than one-quarter of the OUIs that produced an inferred
offset of zero 
between the wired MAC and wireless BSSID. 

At least two potential scenarios 
likely explain the root cause of the zero mode in the
distribution.  First, a device may lack a link-layer
identifier suitable to create an \eui \vsix address for a particular
interface, \eg the cellular interface of a hotspot device. In this case, the
MAC address of a different interface is used (the BSSID) to create the \eui
\vsix address; since the network prefixes should differ on each interface, no
address collision will occur. A second cause for an inferred \wtb offset value
of zero is MAC address reuse between the wired and wireless interfaces. Because MAC
addresses allocated from a vendor's OUI are assumed unique~\cite{ieee802}, this
scenario suggests misuse of IEEE-assigned MAC address space.

\begin{figure*}[t]
  \centering
  \begin{subfigure}{.45\textwidth}
    \centering
    \includegraphics[width=.95\columnwidth]{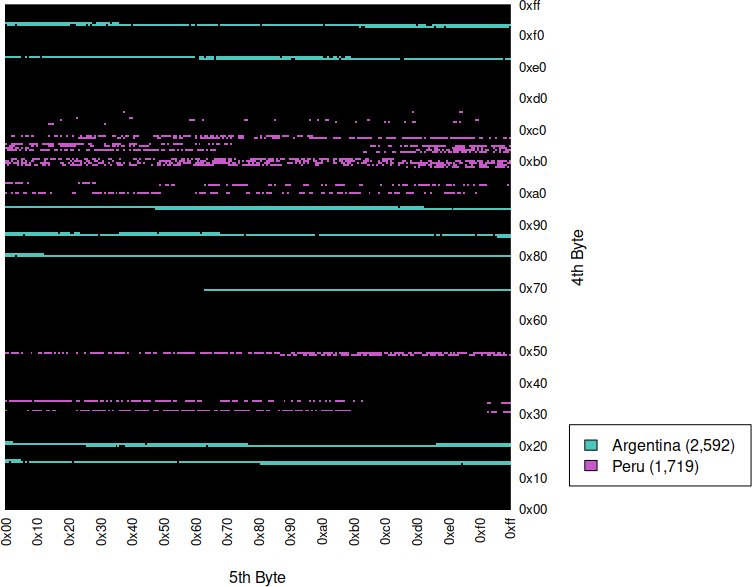}
    \caption{A Mitrastar \ac{OUI} (\texttt{CC:D4:A1}) displays bands of
    MAC address space that geolocate to different, nonadjacent countries.}
    \label{fig:mitrastar}
  \end{subfigure}%
    \hspace{1em}
  \begin{subfigure}{.45\textwidth}
    \centering
      \includegraphics[width=.95\columnwidth]{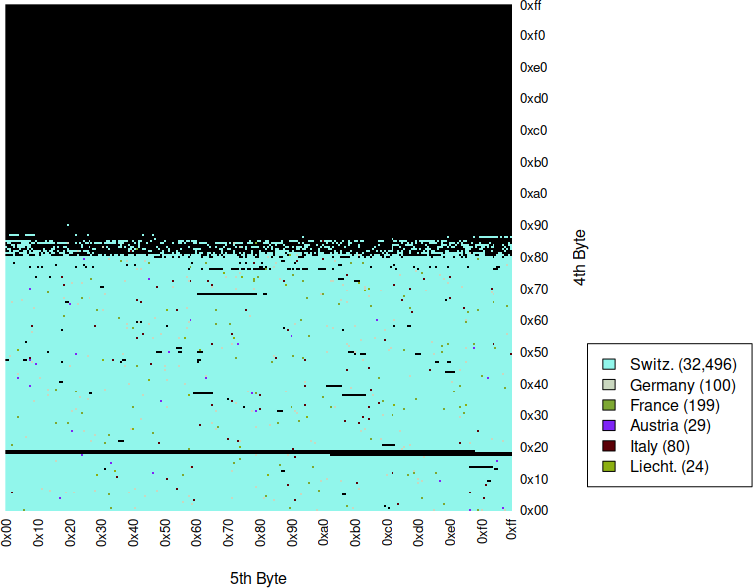}
      \caption{An Askey Corp. \ac{OUI} (\texttt{1C:24:CD}) whose MAC addresses
      are geolocated primarily to a single nation.} %
    \label{fig:askey}
  \end{subfigure}%
  \caption{Inferred country-level geolocation distribution for MAC addresses in
    two \acp{OUI}. Points represent $4^{th}$ ($y$-axis) and $5^{th}$ ($x$-axis) bytes of
    MAC addresses colored according to geolocated country.
}
\label{fig:oui_countries}
\vspace{-2mm}
\end{figure*}

\subsection{Geolocation Inferences}
\label{sec:geocomp}

\begin{table}[b]
    \scriptsize
    \centering
    \caption{\seeu geolocation results; summary of matches between
     \eui-derived WAN MAC addresses and BSSIDs
    from \wifi geolocation databases.}
\begin{adjustbox}{width=1\linewidth}
\begin{tabular}{|p{14mm} l||p{14mm} l|}
\hline
\textbf{Geolocations} & \textbf{Country} & \textbf{Geolocations}
    &\textbf{OUI}  \\ \hline
    3.5M (29.2\%)  & DE        & 603k (5.0\%)  & A0:65:18 (VNPT Tech.) \\ \hline
    1.5M (12.2\%) & US        & 374k (3.1\%)  & 10:86:8C (Arris)  \\ \hline
    1.3M (10.6\%) & VN        & 254k (2.1\%)  & 3C:7A:8A (Arris) \\ \hline
    1.2M (9.6\%) & FR        & 249k (2.1\%)  & A4:F4:C2 (VNPT Tech.) \\ \hline
    1.0M (8.2\%) & BR        & 247k (2.0\%)  & E0:28:6D (AVM GmbH)  \\ \hline
    3.7M (30.3\%) & 142 Other & 10.4M (86\%) & 1,003 Other
    \\ \hline\hline
    \multicolumn{2}{|l||}{12.1M (100\%)} & \multicolumn{2}{|l|}{12.1M
    (100\%)} \\ \hline
\end{tabular}
    \label{tab:results}
\end{adjustbox}
\end{table}

Table~\ref{tab:results} summarizes our results from matching \eui \vsix
address-derived MAC addresses with \wifi BSSIDs. \added{Our algorithm pairs
at least one \vsix address from 1,114 unique ASNs with a geolocated BSSID,
representing approximately 5\% coverage of the $\sim$27k IPv6 ASNs announced in
the global BGP routing table. Of the 347M
unique \eui \vsix addresses in our corpus, 118,429,034 ($\sim$34\%) contain an
embedded  WAN MAC address that pairs with a geolocated BSSID. Further, due to
provider prefix cycling~\cite{rye2021follow} and address churn, multiple \eui
\vsix addresses with the same encoded WAN MAC map to the same BSSID.  Only 12M
unique MAC addresses are encoded in the
118M \eui \vsix addresses that have a WAN MAC-BSSID correlation.}  %

Germany is the most frequently-geolocated country, with more than
a quarter of the total address matches.
This is primarily due to a large number
of \wifi routers made by AVM GmbH under the brand name ``Fritz!Box''.   We note
that, for this reason,  we purchased AVM CPE and explicitly validated \seeu on
these devices -- and thus have high-confidence in the geolocation inferences for
this large subset.

\seeu enables insight into the geographic distribution of
devices within an OUI.  Figure~\ref{fig:oui_countries} displays the
breakdown of country-level geolocations we obtained for \eui \vsix addresses
within two different OUIs.  Figure~\ref{fig:mitrastar}, representing a Mitrastar
OUI, shows that there are distinct bands of MAC addresses allocated to
devices operated in different, non-adjacent South American
countries. Figure~\ref{fig:askey}, on the other hand, shows an Askey Corporation
OUI in which the vast majority of MAC addresses allocated to their devices are
geolocated to a single European country.

\subsection{\eui\vsix Geolocation Comparison}

To further explore \seeu, we compared our geolocation capability with
current widely-used IP geolocation databases.  IP geolocation
databases are known to contain inaccuracies
\eg~\cite{poese2011ip,komosny2017location}, and are generally used for
applications that only require country or city-level accuracy.
However, these databases provide comparative insight into \seeu's
goal of providing street-level geolocation.

\begin{figure}[t]
     \centering
     \includegraphics[width=0.9\columnwidth]{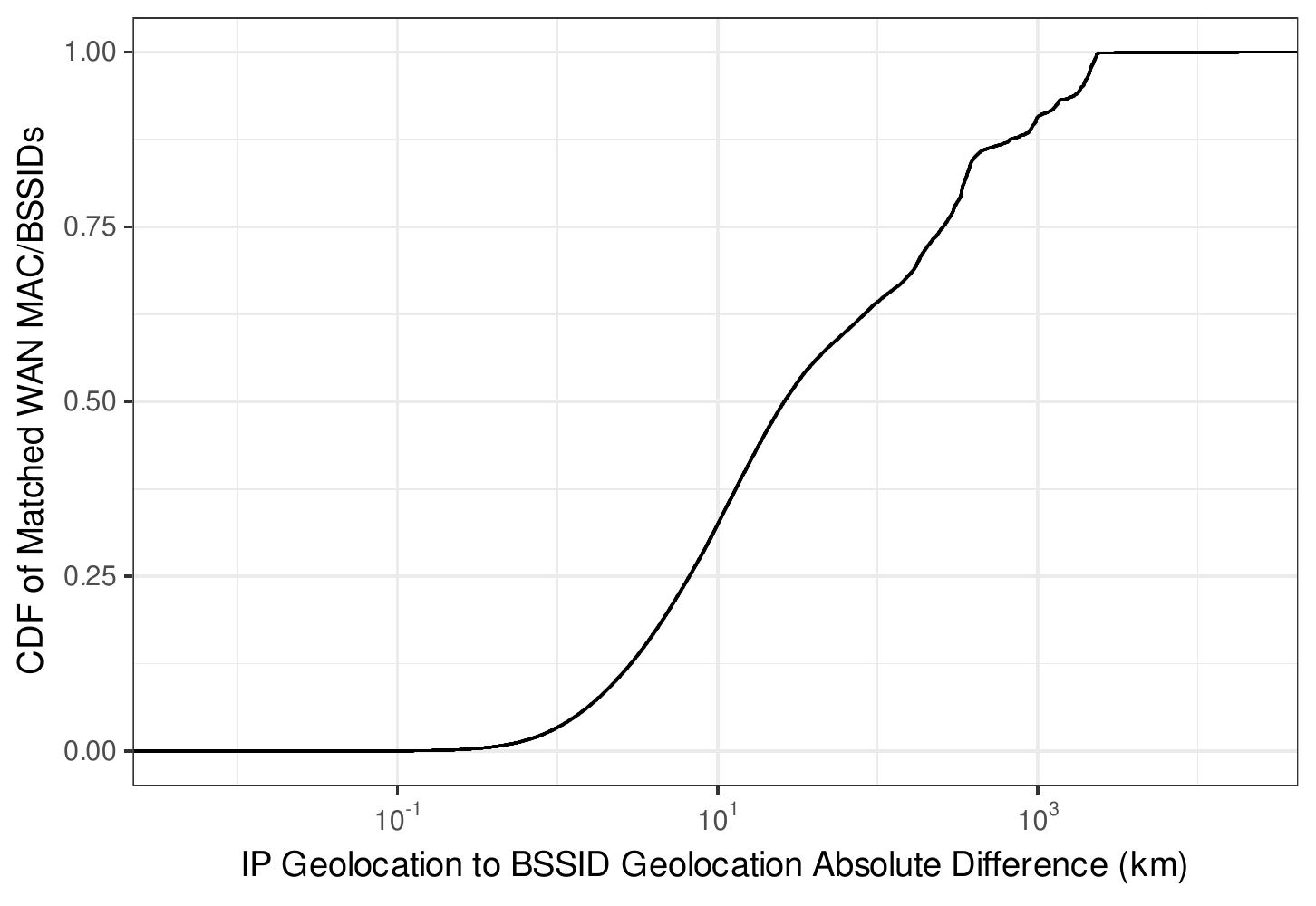}
     \caption{CDF of geolocated \vsix addresses displaying
     the distance difference between \maxm and \seeu.}
     \label{fig:distcdf}
     \vspace{-2mm}
\end{figure}

We first used the popular \maxm GeoLite2
geolocation database~\cite{maxmind} contemporaneous with the active scan data
(April 2021) to
obtain coordinates for the \eui \vsix
addresses in our corpus (\S\ref{sec:euicorpus}).   
For each MAC address in our WAN MAC address corpus, we retrieved the \eui \vsix
address it was embedded in. 
In cases where a MAC address
appeared in more than one \eui \vsix address due to periodic prefix cycling by
ISPs, we randomly selected one of the \vsix addresses for comparison. 
Then, for each of the BSSIDs matched to the WAN MAC addresses present in
our \eui corpus, we compared the BSSID geolocation to the \maxm IP geolocation.
Like the \eui \vsix addresses, BSSIDs appeared in our geolocation data multiple
times with different coordinates as well. %
Again, we chose one of the
geolocations randomly to use as the canonical location. 
Figure~\ref{fig:distcdf} is a CDF of the geodesic distance difference
between \seeu inferences and \maxm geolocations.
Because we do
not know whether \maxm or our inference is correct, this metric is simply
the distance between the two points.  However, we note that when the
wardriving database is accurate and up-to-date, we expect our
inference to better represent the true location.
While approximately 2,800
(0.02\%) WAN MAC-BSSID pairs have \maxm and wardriving geolocations within
100m of each other, about 75\% of all MAC-BSSID pairs have IP and BSSID
geolocations more than 8 kilometers apart. The median difference between \maxm and
our wardriving database locations is 26 kilometers, indicating that the locations
routinely differ on a city- or regional-level. In the extreme case,
geolocations provided by \maxm and our data differ by 
thousands of kilometers. %
There are several potential reasons for
drastic geolocation differences. First, a router may move between \eui address discovery
and BSSID geolocation. %
Secondly, an incorrect BSSID inference from a WAN
MAC address may erroneously match a %
device in a different geographic region than the correct inference would have. Finally,
the \maxm geolocation data may also be incorrect or stale.

\added{We further analyzed the 20\% of geolocations
with the largest difference between \maxm and \seeu locations
(2,423,991 pairs with geolocation differences $\geq$ 335km). Of
these 2.4M WAN MACs encoded in \vsix addresses, five of the ten most common OUIs belong to 
the Arris Corporation, accounting for 45\% (1,093,184 of 2,423,991).
This company produces one of the consumer-grade routers issued
by a large US ISP, leading to a high degree of OUI and ISP homogeneity among
the most extreme geolocation differences. All but ten of these devices are
geolocated to the same point in a lake in Kansas, USA by \maxm's GeoLite2
(presumably representing a default location),
while \seeu produces 1,093,076 unique geolocations throughout the
provider's coverage area. Another three of the ten
most common OUIs belong to Mitrastar, accounting for $\sim$9\% of the top 20\%
of geolocation discrepancies. Of these 213,465 geolocated devices, 96\% (205,594)
are geolocated to a point in Guanabara Bay, Brazil by \maxm, while \seeu reports 213,439 distinct
locations.}

The number of unique geolocation data points in the wardriving and \maxm
datasets suggests that the wardriving data is closer to
ground truth. Of the 347M unique \vsix \eui addresses in the \vsix corpus, \maxm
returns only 22,676 distinct geolocations, indicating that \maxm places millions
of \vsix addresses at the same positions. In contrast, the wardriving data
we obtained comprises 433M distinct BSSID geolocations of 450M total BSSIDs.
This means that far fewer BSSIDs are geolocated to the same point in our
wardriving data; those that do geolocate to the same point are frequently
sequential, indicating that they are likely the BSSIDs for two different \wifi
frequency bands. Focusing specifically on our 12.1M MAC to BSSID matches, \maxm
returns 10,133 distinct \vsix geolocations, while \seeu returns 12.1M.

\begin{figure}[t]
  \centering
  \resizebox{0.9\columnwidth}{0.6\columnwidth}{\includegraphics{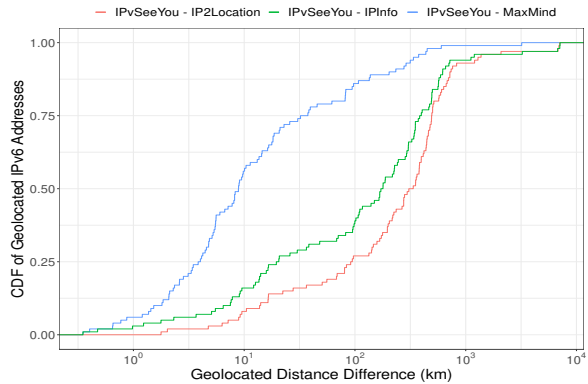}}
    \caption{CDF of geolocation distance differences between \seeu and popular IP
    geolocation databases.}
\label{fig:multi-geo}
    \vspace{-4mm}
\end{figure}

Finally, we compared \seeu's geolocation across other popular IP geolocation services,
including \maxm, IP2Location~\cite{hexasoft}, and IPinfo.io\cite{ipinfo} by
sampling a random 100 geolocated addresses from our corpus and comparing the
resulting geolocation from each service. Although none of these sources of data
are ground truth, Figure~\ref{fig:multi-geo} shows that
\seeu is most consistent with \maxm, with a median geolocation distance
difference of about 10 kilometers. IPInfo and IP2Location have much larger geolocation
differences, with median differences of approximately 250 and 500 kilometers from \seeu,
respectively.

\section{Infrastructure and Non-\eui Geolocation}
\label{sec:infra}

Thus far, our \vsix geolocation capability applies only to \cpe
devices that use \eui addressing and the prefixes associated with
those devices.  In this section, we extend our technique to permit
more general geolocation of \cpe devices via association with
geolocatable \cpe.  While extending our 
coverage comes at the \deleted{potential} cost of reduced accuracy, it
allows for geolocation of devices with unknown offsets, missing
BSSID in the available wardriving databases, and CPE that do
not use \eui addressing. 
In typical deployments, multiple \cpe devices connect to, and are
aggregated by, an upstream provider router.  Further, the network
link between a \cpe and its upstream router is generally relatively
short due to protocol specifications and physical constraints.  For
example, the DOCSIS standard for cable modems defines a maximum
distance between the CMTS and modem of 100 miles (160 kilometers),
but with a ``typical maximum separation of 10-15 miles.''
\cite{docsis}  

We therefore leverage \seeu-geolocated \eui \cpe to locate: 1) upstream provider last mile
infrastructure; 2) \eui \cpe that we cannot geolocate using our methodology in
\S\ref{sec:matching}; and 3) non-\eui \cpe.
Our basic intuition is straight-forward: known locations of \cpe
devices can be used to infer the location of unknown \cpe if they
connect to the same provider router (and, hence, are likely in close 
physical proximity).  
Further, when our assumptions
about the distance between a \cpe and the router to which it connects
are incorrect, \eg in a virtualized network topology, this error will be
reflected by a wide dispersion of geolocated devices and thus evident 
and detectable. 

\begin{figure*}[t]
  \begin{subfigure}{.5\textwidth}
    \centering
    \includegraphics[width=.75\columnwidth]{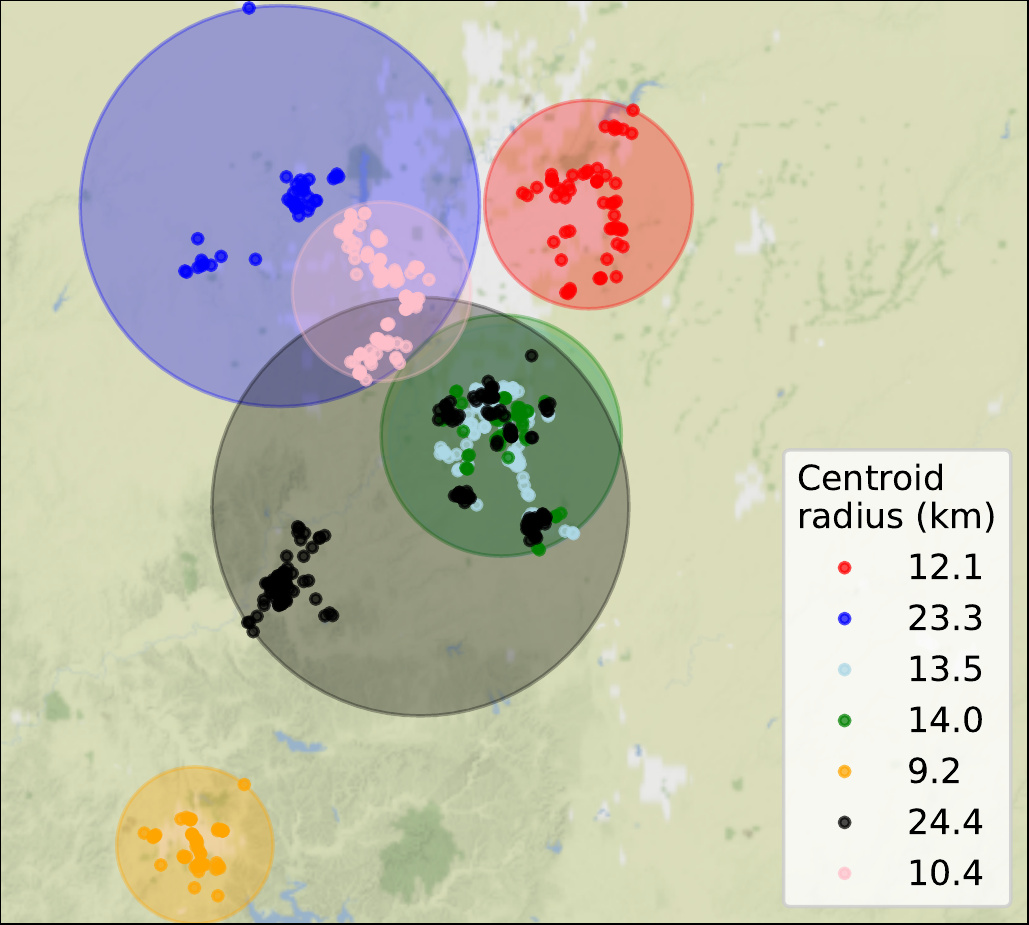}
  \caption{Indianapolis, IN, USA \cpe}
    \label{fig:indy-infra}
  \end{subfigure}%
  \begin{subfigure}{.5\textwidth}
    \centering
      \includegraphics[width=.75\textwidth]{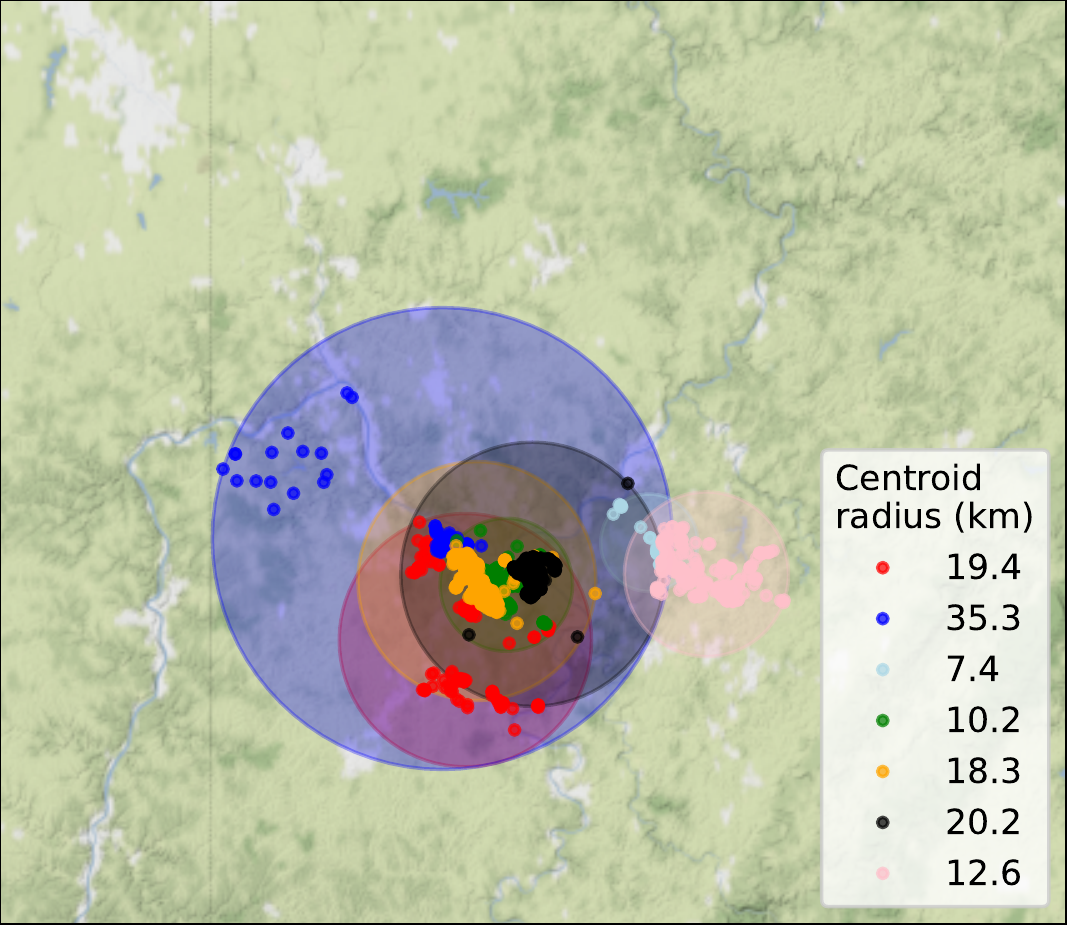}
      \caption{Pittsburgh, PA, USA \cpe}
    \label{fig:pitt-infra}
  \end{subfigure}%
    \caption{\cpe geolocations for two large US
metropolitan areas; colors represent a common penultimate
(provider) router.  Non-\eui \cpe that connect to the same
router are inferred to be within the same cluster as their
\eui counterparts.}
\label{fig:infra}
\end{figure*}

\subsection{Infrastructure Case Study}

To explore the feasibility of this intuition, we probed the path toward
each prefix behind all \eui routers in our corpus that are connected
to a large United States
residential ISP. 
For this probing, we used \yarrp
from a well-provisioned vantage point;
\yarrp
is a high-speed randomized IP topology prober that reveals
the sequence of router interfaces along the data plane path in the
same fashion as traceroute~\cite{imc16yarrp,imc18beholder}.  
We then grouped successfully geolocated \eui \vsix
\cpe by their penultimate hop, \ie the CPE's upstream provider router.  

Figure~\ref{fig:infra} depicts the geolocation of \eui \cpe mapped by
\seeu in two 
metropolitan areas of the
United States.  Each \cpe is a dot on the map, while each color
corresponds to a common penultimate hop (provider router) 
revealed via \yarrp. For each 
infrastructure router, we additionally compute the centroid of the set
of geolocated
\cpe it serves. Then, we plot a lighter circle of the ISP router's color
around the centroid with minimum radius $r$ such that all geolocated \cpe fall
within $r$ kilometers. This illuminates an
inferred coverage area for each provider router.

As an illustrative example showing both the power of the technique as
well as the complexity of real-world deployments, Figure~\ref{fig:indy-infra} maps 
seven distinct clusters in
Indianapolis, IN, USA corresponding to seven different provider routers. 
Fifty-seven \cpe geolocate in the red cluster and 173 \cpe geolocate to
the black cluster. The red \cpe form a fairly dense grouping, with all \cpe
within 12.1 kilometers of the geolocations' centroid. The black \cpe, however, are
distributed between two distinct geographic clusters approximately 10 kilometers apart. Further,
the northeastern grouping in the black cluster substantially overlaps with both the green and light
blue clusters. For candidate \cpe that either do not use \eui addressing or
that
we fail to geolocate directly using \seeu, infrastructure router  
clustering provides an indirect, coarse geolocation mechanism.

Figure~\ref{fig:pitt-infra} shows a more complicated, yet equally
compelling, example from the Pittsburgh, PA, USA area.  Here, there
are again seven different colors representing seven unique provider
routers for the geolocated \cpe. %
In this example, significant overlap exists in the provider router service
areas. All seven penultimate hops have at least one \cpe device located inside
of the dark blue coverage area, and the red, orange, green, and black coverage
areas substantially intersect. This result is expected, as multiple ISP
infrastructure routers may be necessary to support deployments in dense
metropolitan regions.

Due to our ability to deduce the service coverage range of a provider's
last-mile infrastructure, even a single device using \eui addressing can
potentially compromise the geolocation privacy of \emph{all} of the devices that connect to
the same infrastructure.  A \cpe using random \vsix addresses
is therefore not sufficiently
protected -- simply living near an \vsix \eui \cpe device can be a privacy
vulnerability.  This further implies that, even if
\eui becomes less widely deployed, legacy equipment that is
infrequency updated or refreshed implementing \eui
will continue to enable geolocation.

\subsection{Accuracy}

\subsubsection{Volunteers}
\label{sec:noneuivols}

\begin{figure}[t]
  \includegraphics[width=\columnwidth]{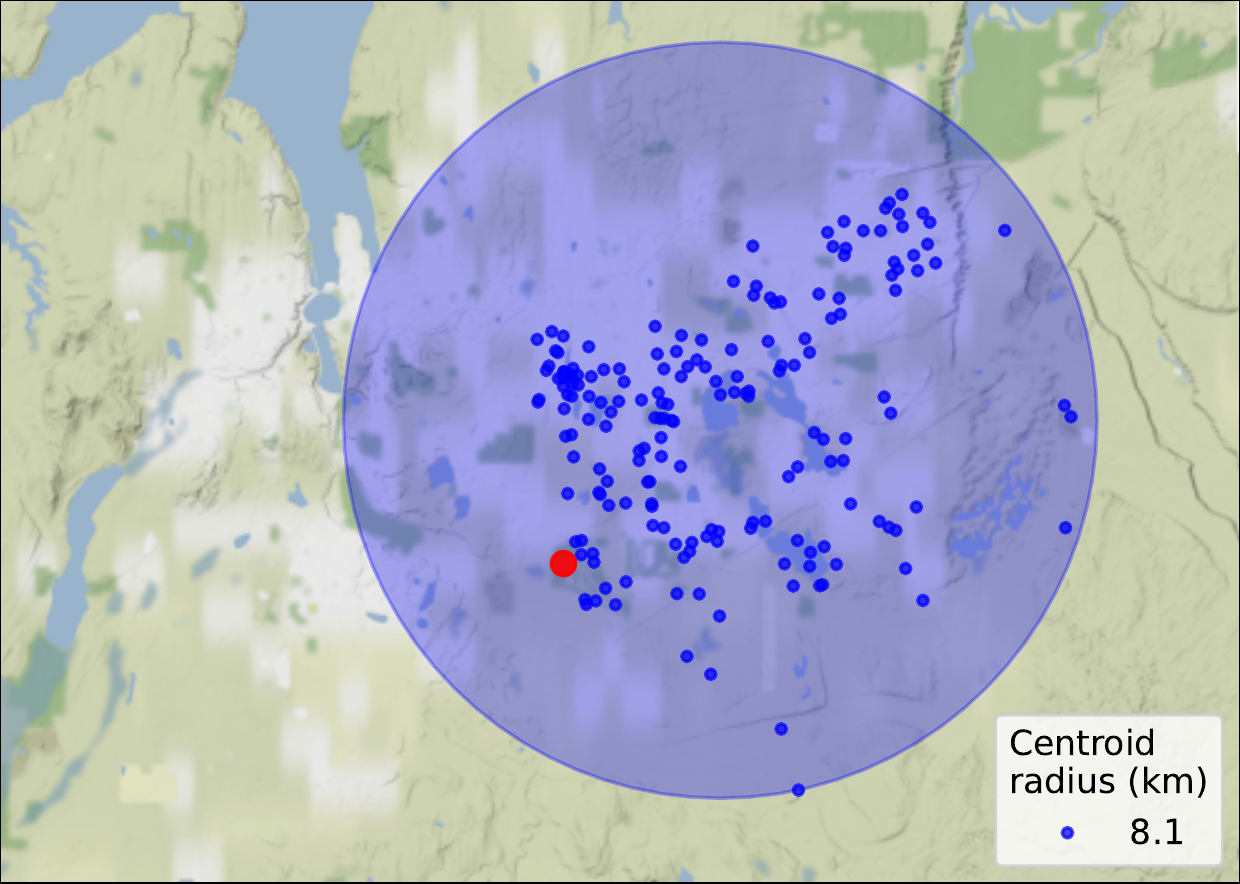}
      \caption{Olympia, WA, USA geolocated \cpe (blue) used to infer the
      location of a known ground-truth device (red) with a non-\eui \vsix
      address.}
\label{fig:olygt}
    \vspace{-4mm}
\end{figure}

\begin{figure}[t]
  \includegraphics[width=\columnwidth]{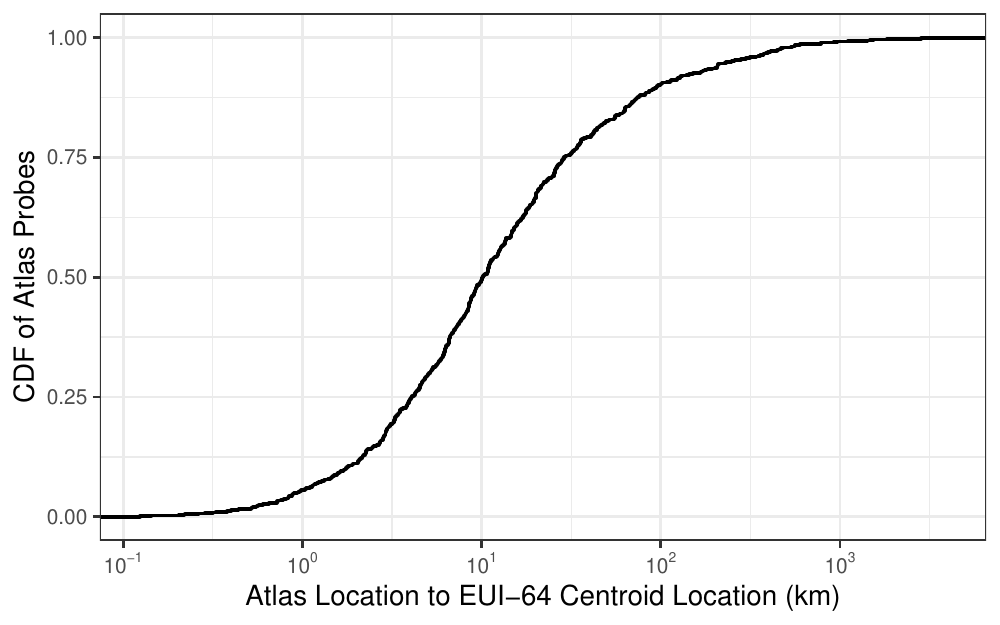}
      \caption{CDF of RIPE Atlas probes displaying the distance between the reported
      probe location and \seeu-derived location.}
\label{fig:atlascdf}
    \vspace{-3mm}
\end{figure}

As in~\ref{sec:euivols}, we solicited volunteers to assist with
validating our methodology for geolocating non-\eui \cpe. The participants
disclosed their LAN \vsix subnet, which we then used to obtain their
\cpe WAN \vsix address via \yarrp
and to discover other \cpe devices assigned adjacent subnet allocations. The nearby
\cpe devices with \eui addresses were then used to infer the BSSID assigned to
the same device using our offset inferences computed in \S\ref{sec:offset},
which in turn were used for precise geolocation using \seeu.

Figure~\ref{fig:olygt} presents geolocated \eui \cpe in blue located
nearby one volunteer's
non-\eui \cpe device (represented as the red point) for which we
have known ground-truth
location (in Olympia, WA, USA). The ground-truth device is located
4.75km from the
\seeu inferred centroid of the geolocated \eui \cpe, demonstrating in
this example that our non-\eui
geolocation methodology produces correct and highly accurate results.

Five additional volunteers from \S\ref{sec:euivols} whose \cpe could not be
geolocated directly using \seeu were geolocated using \seeu on \eui \cpe in
adjacent subnet allocations. These non-\eui devices' true locations all fall within
between 550 meters to 9 kilometers of the centroids of the \eui \cpe allocated
prefixes adjacent to our non-\eui ground truth. 

\subsubsection{RIPE Atlas}
\label{sec:atlasvalidation}

For additional evaluation of \seeu's ability to geolocate non-\eui
\cpe,  
we utilize RIPE
Atlas~\cite{atlas} ``probes.''  Probes are lightweight
measurement nodes installed in homes and networks.
Currently, Atlas has approximately 25,000
probes distributed throughout the world~\cite{atlas}.
Probe owners self-report their device's physical coordinates into the RIPE Atlas
database when registering. When Atlas data is queried by non-owners,
RIPE inserts an error of up to one kilometer to preserve
the owner's privacy~\cite{ripeerror}. 
While some probe owners may intentionally input incorrect
geolocation coordinates, we assume that most users disclose a reasonably accurate device
location to RIPE and examine the data in aggregate.  

We consider only RIPE probes with \vsix connectivity, and those
in residential networks.
We first eliminate RIPE probes that indicate they are in a
data center or use a known tunnel broker \vsix prefix,
which reduces the total number of probes we examine to approximately
3,500. Then, we initiate \yarrp traces to address space adjacent to
the Atlas probes
to elicit responses from nearby \cpe routers. For 893 probes, we
obtain at least one \eui address in the same /48 prefix as the probe device 
we are attempting to geolocate.  In this experiment, all CPE in the
same /48 have the same penultimate infrastructure hop (provider
router).  For a given probe, we use \seeu to geolocate
the \eui CPE attached to the same provider router (in the same
/48) as the probe.  Then, we find the centroid of the associated \eui
CPE geolocations and use it as the inferred geolocation of the probe.

Figure~\ref{fig:atlascdf} displays the error between the \seeu 
inferred Atlas probe locations 
and the reported
location of the probe. The latter 
includes
both the RIPE-injected error and any error introduced by the device owner when
self-reporting their probe location.
Nonetheless, the median distance is approximately ten
kilometers, indicating that our methodology consistently geolocates
the large and widely distributed set of RIPE probes
with high accuracy. In some instances, our methodology detects probes that
have likely changed locations without an accompanying RIPE update. For instance, one probe
was purportedly in Los Angeles, CA, USA, but 
all other \eui \cpe in the same /48 as the probe geolocate to 
a Seattle, WA, USA suburb. 

\begin{figure}[t]
  \includegraphics[width=0.9\columnwidth]{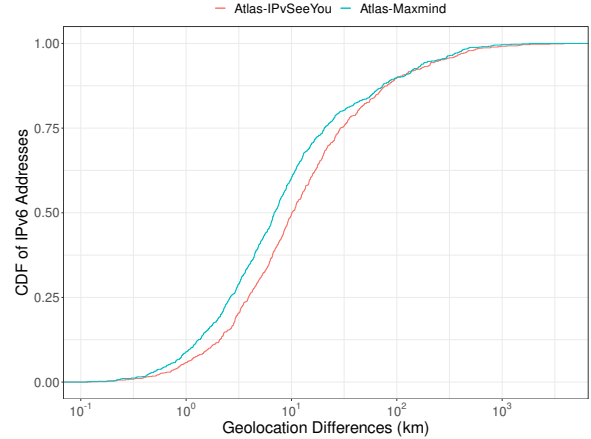}
    \caption{A CDF of RIPE Atlas nodes (non-\eui) depicting the distance difference
    between the reported location and \maxm and \seeu-centroid
    geolocations.}
\label{fig:atlasmaxmind}
    \vspace{-2mm}
\end{figure}

Finally, we compare our inferred centroid locations to \maxm's IP geolocations.
Figure\ref{fig:atlasmaxmind} displays the difference between the RIPE-reported
location (including the error) and the \maxm and \seeu-centroid geolocations. Note that
none of these geolocations \added{(\maxm, \seeu-centroid, or RIPE)} represents
ground truth, and, in the case of the \seeu-centroid,
the geolocation is an aggregate of other nearby \cpe geolocations. However, 
there is general agreement between RIPE Atlas' \added{error-injected} reported location,
\maxm, and \seeu. \added{We claim only feasibility, rather than improved
accuracy over commercial databases, for non-\eui/infrastructure geolocation.}

\vspace{-2mm}
\subsection{Coverage Gain from \eui Clustering}

Using our technique for associating clusters of geolocated \eui \cpe with
non-geolocatable  \eui and non-\eui \cpe, we last 
evaluate the coverage gain.
The coverage gain is
simply the net increase in additional \cpe that can be geolocated by
leveraging the locations of other \cpe connected to the same provider
router. As an example, 
we
perform \yarrp traces to random \acp{IID} in each /64 within a single
/48 of the large American ISP previously considered.
In this exemplar /48, we discover 3,825 distinct \cpe addresses, including 1,776
(46\%) \eui and 2,049 \vsix addresses with random IIDs.
Because /60 subnets are allocated to end-users in this /48, we would expect to
see at most 4,096 unique \vsix \cpe addresses; thus, the /48 is nearly
completely allocated and discoverable.

Employing \seeu from \S\ref{sec:methodology}, we geolocate 180 of the
1,776 \eui \cpe in the service provider's /48 to the Olympia,
Washington, USA
metropolitan region, displayed in Figure~\ref{fig:olygt}. Assuming that the
distribution of non-geolocatable \eui and non-\eui \cpe does not vary
significantly from the 180 geolocated \cpe, we should expect to find these
``hidden'' devices in approximately the same location as the 180 geolocated
\cpe. Thus, our methodology enables the geolocation of all 3,825 \cpe as opposed
to the 180 for which we have precise geolocations, representing a CPE
coverage gain 
factor of approximately 20.
\section{Remediation}
\label{sec:remediation}

Despite privacy-preserving mechanisms for dynamically generating \vsix addresses
existing for two decades~\cite{rfc3041}, tens of millions of CPE devices continue 
to use \eui \vsix addresses. \eui \ac{SLAAC} addressing,
when combined with predictable MAC address assignments
to device interfaces, enables an adversary to conduct the type of data fusion
attack we outline in \S\ref{sec:methodology} and demonstrated the feasibility
of in \S\ref{sec:results}. The most straightforward solution to our attack is
for more CPE vendors to employ a random addressing
mechanism~\cite{rfc4941,rfc7217,rfc8981} to generate its WAN 
\vsix address. %

Toward the goal of encouraging vendors to adopt countermeasures to our attack, we
responsibly disclosed our findings to the manufacturers of devices tied to over
7M \eui-derived MAC addresses in our corpus. %
Based on our findings, one
manufacturer plans to transition to using modern randomized \acp{IID} when the device
generates WAN \vsix addresses.  A second vendor
disputed our findings, specifically that their devices were exposing
MAC addresses via \eui, despite clear evidence to the contrary.  %

Several additional mitigations for our attack also exist. CPE
manufacturers can protect their devices from our data fusion attack by
randomizing their MAC address allocation patterns. This breaks
the linkage we infer between the \eui-derived WAN MAC address and
BSSID.
This mitigation requires the vendor to
ensure that no fixed patterns exist in MAC address allocation, and
maintain a strict account of which randomly-chosen addresses have already been
allocated to avoid duplicate MAC assignment. Furthermore, continued use of \eui
SLAAC addressing still permits targeted attacks and device tracking.
Therefore, we view this as a suboptimal solution.

A third protection mechanism against our attack is the use of randomized MAC
addresses for either or both of the WAN MAC and 802.11 BSSID. %
Adopting MAC address randomization on either the WAN or 802.11
interface, wherein a new, random MAC address is generated at each power cycle or
when a new \eui \vsix address is generated, would also prevent linking the two
identifiers, and thus the geolocation fusion attack. However, the technical
difficulty of implementing MAC address randomization in CPE devices is likely to
be as or more difficult than simply enabling random %
addressing. Further, a CPE device with BSSIDs that change over time might be
erroneously construed as an attacker conducting an ``Evil-Twin''
attack~\cite{lanze2015hacker,bauer2008mitigating,vanhoef2016mac}.

A final protection mechanism is to disable ICMPv6 responses on the CPE.
This prevents an attacker from obtaining the CPE WAN address
through active scans. However, in disabling ICMPv6, the operator also loses the
ability to use ICMPv6 responses to troubleshoot networking issues. In
many low-cost CPE, no mechanism to disable ICMPv6 is even exposed. %

We strongly suggest the use of random \vsix addresses due to the limitations of 
other potential mitigations. However, as noted in \S\ref{sec:infra}, unless
\emph{all} \cpe devices employ these mitigations, even a single \eui device can
aid an attacker in geolocating non-\eui \cpe devices. 
\section{Summary and Conclusions}
\label{sec:conclusion}

In this work, we demonstrated a location privacy vulnerability that exists in
millions of deployed \vsix devices. Despite best current practices discouraging
the use of \eui \vsix addresses and their disuse in most modern endpoint
operating systems, many CPE devices continue to generate \vsix addresses by
embedding the interface's MAC address in the lower 64 bits of the \vsix address.
Further, many CPE manufacturers assign MAC addresses predictably to the
interfaces on their devices. Due to these two design choices, we were able to
fuse two large datasets, one consisting of \eui-derived MAC addresses, the other
of \wifi BSSID geolocations, to correlate these identifiers and discover the
physical location of millions logical \vsix addresses.

Toward this end, we developed a novel algorithm to determine the number of MAC
addresses assigned to individual CPE devices and infer the offset between the
WAN interface MAC address and \wifi BSSID.  We found that over 12M \eui \vsix
addresses in 146 countries could be matched to \wifi BSSIDs.
Not only does this privacy
vulnerability impact device owners whose products implement this legacy
technology, but nearby devices connected to the same provider router can also be
geolocated due simply to their proximity to \eui \cpe. The insecurity of even a
few legacy devices jeopardizes the privacy of all of their neighbors, no matter
how privacy-conscious.

Due to the consequences of our location privacy attack, we contacted several CPE
vendors as well as a large ISP.
Our results lead to the
deprecation of \eui addressing by a one manufacturer, and mitigation
of the vulnerability within the network of the large residential
service provider.

However, residential routers are rarely updated and infrequently
replaced.  Thus, the ossified ecosystem of deployed residential cable and DSL
routers implies that \seeu will remain a privacy threat into the
foreseeable future.
 
\section*{Acknowledgments}
We thank kc claffy, Oliver Gasser, Will van Gulik, 
Thomas Krenc, Josh Kroll,  Jeremy Martin, Justin Rohrer,
and the anonymous reviewers
for invaluable feedback, bobzilla from WiGLE, and Kevin Borgolte for shepherding.
Authors Rye and Beverly are founders of sixint.io, which
specializes in IPv6 intelligence and security solutions.

\bibliographystyle{IEEEtran}
\bibliography{edgy}

\begin{acronym}
  \acro{AS}{Autonomous System}
  \acro{AP}{Access Point}
  \acro{CAIDA}{Center for Analysis of Internet Data}
  \acro{CPE}{Customer Premises Equipment}
  \acro{EUI}{Extended Unique Identifier}
  \acro{FPGA}{Field-Programmable Gate Array}
  \acro{FCC}{Federal Communications Commission}
  \acro{IID}{Interface Identifier}
  \acro{ISP}{Internet Service Provider}
  \acro{MAC}{Media Access Control}
  \acro{NAT}{Network Address Translation}
  \acro{OUI}{Organizationally Unique Identifier}
  \acro{SOHO}{Small Office-Home Office}
  \acro{SLAAC}{Stateless Address Autoconfiguration}
  \acro{WiGLE}{Wireless Geographic Logging Engine}
  \acro{WAN}{Wide Area Network}
\end{acronym}

\clearpage
\newpage
\section*{Appendix: WAN and WiFi Corpora}

\begin{table*}[t]
    \centering
    \caption{Summary of top countries, ASes, and OUIs of MACs embedded in \eui
    \vsix addresses. MAC addresses found in more than one \ac{AS} are not included to
    account for potential MAC address reuse.}
    \label{tab:topeuistats}
\begin{adjustbox}{max width=\textwidth}
\begin{tabular}{|l r||l r||l r|}
\hline
\textbf{Country} & \textbf{Count} &  \textbf{AS} & \textbf{Count} &
\textbf{OUI / Manufacturer} & \textbf{Count} \\    \hline
    CN & 21,425,581 (35.4\%) & Comcast (AS7922) & 10,188,218 (16.8\%) &
    14:AD:CA / China Mobile IOT & 904,783 (1.5\%) \\ \hline
    US & 11,196,587 (18.5\%) & Guangdong Mobile (AS9808) & 8,004,879 (13.2\%) &
    F0:3C:91 / Unknown          & 885,386 (1.5\%) \\ \hline
    DE & 9,265,924 (15.3\%)  &  Deutsche Telekom (AS3320) & 6,353,101 (10.5\%) & B0:30:55
    / China Mobile IOT & 875,657 (1.4\%) \\ \hline
    BR & 3,404,573 (5.6\%)   & France Telecom (AS3215) & 2,746,829 (4.5\%)  & FC:8E:5B /
    China Mobile IOT & 839,804 (1.4\%) \\ \hline
    FR & 2,753,927 (4.5\%)   & China Unicom (AS4837)    & 2,399,925 (4.0\%)  & FC:F2:9F /
    China Mobile IOT & 738,947 (1.2\%) \\ \hline
    195 Other & 12,525,250 (20.7\%)& 12,651 Other & 30,878,890 (51.0\%) &
    463,183 Other &
    56,327,265 (93.0\%) \\ \hline
\end{tabular}
\end{adjustbox}
\end{table*}

\begin{table*}[t]
    \centering
    \caption{Summary of BSSID geolocation dataset by
    geolocated country, BSSID manufacturer, and data source. The total number of
    unique BSSIDs is less than the sum of the individual data sources due to
    some BSSIDs existing in multiple datasets. A small number of BSSIDs
    (particularly small constants such as \texttt{00:00:00:00:00:01}) geolocate
    to multiple countries.}
    \label{tab:geostats}
\begin{adjustbox}{max width=\textwidth}
\begin{tabular}{|l r||l r||l r|}
\hline
    \textbf{Country}   & \textbf{Count}        & \textbf{OUI / Manufacturer} &
\textbf{Count} & \textbf{Source} & \textbf{Count}                        \\ \hline
    US & 119,591,390 (26.6\%) & A0:65:18 / VNPT Technology     & 2,206,621 (0.5\%) & Apple API & 444,860,460 \\ \hline
    DE & 78,034,169	(17.3\%) & 98:9B:CB / AVM GmbH    & 1,352,222 (0.3\%) & OpenWifi.su & 29,340,881 \\ \hline
    BR & 37,245,817  (8.3\%) & 3C:A6:2F / AVM GmbH    & 1,320,865 (0.3\%) & Mylnikov & 20,226,869 \\ \hline
    FR & 32,464,391 (7.2\%) & 7C:FF:4D / AVM GmbH    & 1,311,865 (0.3\%) & OpenBMap & 15,384,623 \\ \hline
    JP & 28,170,359  (6.3\%) & 38:10:D5 / AVM GmbH    & 1,282,799 (0.3\%) & WiGLE & 1,367,700 \\ \hline
    233 Other & 154,588,509 (34.3\%) &  850,083 Other & 442,543,751 (98.3\%) & Total & 450,018,123 \\ \hline
\end{tabular}
\end{adjustbox}
\end{table*}

Because the same MAC address can appear in many \eui \vsix addresses, we
characterize our data using the MAC address as the primary unit of
measure. For example, Versatel 1\&1's (AS8881) prefix delegation policy causes
many \cpe devices to generate a new \eui addresses every 24
hours~\cite{rye2021follow}; in other providers, the delegated prefix and \cpe
\eui address may remain stable for several
months~\cite{padmanabhan2020dynamips}. \added{In our data, the 
maximum number of \vsix addresses to a single WAN MAC in our corpus is 5,652,
with a mean of 9.8 and median of 1 \vsix addresses per WAN MAC. Note that these
statistics depend substantially on the networks
probed and the duration of the probing.}

The 60,571,842 \added{WAN} MAC addresses \added{in our corpus} are embedded in \eui \vsix addresses from
\acp{AS} corresponding to 200 different countries and territories as determined by Team Cymru's
IP-to-ASN lookup service~\cite{cymru2008ip}. China contributes the largest fraction at
35\%; the top ten countries each add over 1 million MAC addresses to the total.
Although China leads all countries in MAC address count, Comcast, an
American \ac{ISP}, is the top \ac{AS} with over 10 million distinct MAC
addresses. Comcast dominates the US contribution
with 91\% of the US \eui-derived MAC addresses; Guangdong Mobile, the leading Chinese AS,
contributes only 37\% of the total Chinese MAC addresses by contrast. 

\added{The OUI and manufacturer data we collect indicate that we discover
463,188 distinct OUIs embedded in \eui \vsix addresses. However, only 32,345
OUIs are listed in a recent IEEE OUI database~\cite{oui}. This discrepancy has
several potential root causes.}

\added{One basis for this variation is due to the
randomness involved in generating temporary~\cite{rfc4941,rfc8981} or
stable~\cite{rfc7217} random addresses. Because \eui \vsix addresses are
identified through a static \texttt{0xFFFE} in the fourth and fifth byte
positions of the IID, a random process would be expected to produce
\emph{false-\eui} IIDs with probability $p= \frac{1}{65,536}$ (or $p =
\frac{1}{131,072}$ if we require the U/L bit to be set). This type of
falsely-identified MAC address is highly unlikely to result in a MAC-BSSID
offset inference and IP-BSSID geolocation because there are unlikely to be any
BSSIDs in the false MAC address' OUI.}

\added{A second underlying cause for the inflated OUI count is due to
\emph{real} \eui \vsix addresses being formed from a MAC address whose OUI is
not registered in the IEEE OUI database. We observe ample evidence of
unregistered OUIs in \eui \vsix addresses in our corpus.  Four of the top five
OUIs we observe resolve to the China Mobile IOT Company; the fifth
(\texttt{F0:3C:91}) is not listed among IEEE-registered OUIs. However, all instances
of MAC addresses using this OUI originate in \eui \vsix addresses from
an American cloud hosting provider's networks (Linode).
Table~\ref{tab:topeuistats} summarizes our WAN MAC address data derived from
\eui \vsix addresses. 
}

\added{Table~\ref{tab:geostats} displays our BSSID geolocation \ds discussed
in \S\ref{sec:wificorpus}. The most commonly-geolocated country for our BSSIDs
is the US, followed by Germany, Brazil, France, and Japan. Of note, China, which
is the most-common country from our WAN MAC \ds, ranks 32 in most-common
BSSID geolocations.}

\added{As with our \eui-derived MAC address \ds, we observe a significantly
larger number of OUIs in our BSSID data than exist in the IEEE OUI database
(850,083 vs 32,345). Again, several root causes for this discrepancy are
possible.} 

\added{First, many \acp{AP} will invert the U/L bit
of their BSSID to form virtual
WLANs using a single NIC. We see evidence of this occurring; for instance, in
our data we observe the TP-Link OUI \texttt{C0:4A:00} occur nearly 379k times in
our BSSID corpus. This OUI with the U/L bit inverted (\texttt{C2:4A:00}) also
appears in our data in 3,108 BSSIDs.} 

\added{Other potential causes include users spoofing \ac{AP} BSSIDs from unassigned
OUI space, or wireless client addresses being erroneously uploaded to
geolocation databases as \ac{AP} BSSIDs. Because probe requests are typically
sent from ephemeral, random source MAC addresses in modern mobile operating
systems~\cite{fenske2021three,vanhoef2016mac,martin2017study}, probe requests
entering the geolocation corpus could potentially add a large number OUIs to the
BSSID corpus.} 

\added{Nonetheless, the majority of our BSSID data come from IEEE-assigned
OUIs. Over $75\%$ (333,996,812 of 442,543,751) of unique BSSIDs come from allocated
OUI space, while $\sim$89\% (392,926,586 of 442,543,751) of OUIs are from
allocated OUI space or allocated OUIs with the U/L bit inverted.}

\end{document}